\documentclass[preprint2]{aastex}

\shorttitle{Magnetically Supported Two-Temperature Disks}
\shortauthors{Oda et al.}

\begin{document}


\title{Thermal Equilibria of Optically Thin, Magnetically Supported,
Two-Temperature, Black Hole Accretion Disks}


\author{H. Oda\altaffilmark{1,4}, M. Machida\altaffilmark{2},
  K.E. Nakamura\altaffilmark{3} and R. Matsumoto\altaffilmark{4}}


\altaffiltext{1}{Harvard-Smithsonian Center for Astrophysics, 60 Garden
Street, Cambridge, MA 02138, USA; hoda@cfa.harvard.edu}
\altaffiltext{2}{Department of Physics and Astrophysics, Nagoya
University, Furo-cho, Chikusa-ku, Nagoya, Aichi 464-8602, Japan}
\altaffiltext{3}{Department of Sciences, Matsue National College of
  Technology, 14-4 Nishiikuma-cho, Matsue, Shimane 690-8515, Japan}
\altaffiltext{4}{Department of Physics, Graduate School of Science,
  Chiba University, 1-33 Yayoi-cho, Inage-ku, Chiba 263-8522, Japan}


\begin{abstract}
We obtained thermal equilibrium solutions for optically thin,
 two-temperature black hole accretion disks incorporating magnetic
 fields. The main objective of this study is to explain the bright/hard state
 observed during the bright/slow transition of galactic black hole
 candidates. We assume that the energy transfer from ions to electrons
 occurs via Coulomb collisions. Bremsstrahlung, synchrotron, and
 inverse Compton scattering are considered as the radiative cooling
 processes. In order to complete the set of basic equations, we specify
 the magnetic flux advection rate instead of $\beta = p_{\rm gas}/p_{\rm
 mag}$. We find magnetically supported (low-$\beta$), thermally stable
 solutions. In these solutions, the total amount of the heating via the
 dissipation of turbulent magnetic fields goes into electrons and
 balances the radiative cooling. The low-$\beta$ solutions extend to
 high mass accretion rates ($\gtrsim \alpha^{2} {\dot M}_{\rm Edd}$) and
 the electron temperature is moderately cool ($T_{\rm e} \sim 10^{8} -
 10^{9.5} {\rm K}$). High luminosities ($ \gtrsim 0.1 L_{\rm Edd}$) and
 moderately high energy cutoffs in the X-ray spectrum ($ \sim 50 - 200 ~
 {\rm keV}$) observed in the bright/hard state can be explained by the
 low-$\beta$ solutions. 
\end{abstract}





\section{Introduction}
Galactic black hole candidates (BHCs) are known to exhibit transitions
between different X-ray spectral states. 
Typically, a transient outburst begins in the low/hard
state at a low luminosity. The X-ray spectrum in the low/hard state is
roughly described by a hard power law with a high energy cutoff at $\sim
200 ~ {\rm keV}$. As the luminosity increases, these systems 
undergo a transition to the high/soft state (so-called a hard-to-soft
transition). The X-ray spectrum in the
high/soft state is dominated by the disk emission of characteristic
temperature $\sim 1 ~ {\rm keV}$.

Recently, two distinct types of hard-to-soft transitions, the
bright/slow transition and the dark/fast transition, are reported
\citep[e.g.,][]{bell06,gier06}. The bright/slow transition occurs at
$\sim 0.3 ~ L_{\rm Edd}$ and takes more than $30$ days. The system 
undergoes a transition from the low/hard state to the high/soft state via
the ``bright/hard'' state and the very high/steep power law (VH/SPL)
state during the bright/slow transition. The X-ray spectrum in the
bright/hard state is described by a hard power-law with a (moderately)
high energy cutoff  at $\sim 50 - 200 ~ {\rm keV}$ and the luminosity is
``brighter'' than that in the low/hard state. The
dark/fast transition occurs at less
than $0.1 ~ L_{\rm Edd}$ and takes less than $15$ days. The system 
immediately switches from the low/hard state to
the high/soft state during the dark/fast transition. Here, $L_{\rm
Edd} = 4 \pi c G M / \kappa_{\rm es} \sim 1.47 \times 10^{39} \left(M
/10 M_\sun \right) \left(\kappa_{\rm es} / 0.34 ~ {\rm cm}^2 ~ {\rm
g}^{-1} \right)^{-1} {\rm erg} ~ {\rm s}^{-1}$ is the Eddington
luminosity, $M$ is the black hole mass, and $\kappa_{\rm es}$ is the
electron scattering opacity.

\cite{miya08} analyzed the results of {\it
RXTE} observations of the BHC GX 339-4 in the rising phases of the
transient outbursts (this object showed the bright/slow transition in
the 2002/2003 outburst and the dark/fast transition in the 2004
outburst). They found that the cutoff energy strongly anti-correlates
with the luminosity and decreases from $\sim 200 ~ {\rm keV}$ to $\sim
50 ~ {\rm keV}$ in the bright/hard state, while the cutoff energy is 
roughly constant at $\sim 200 ~ {\rm keV}$ 
in the low/hard state. This suggests that the electron
temperature of an accretion disk emitting hard X-rays decreases as the
luminosity increases in the bright/hard state. Furthermore, the
bright/hard state has been observed from $\sim 0.07 ~ L_{\rm Edd}$ up to
$\sim 0.3 ~ L_{\rm Edd}$. They concluded that such anti-correlation is
explained by the scenario that
the heating rate from protons to electrons via the Coulomb collision
balances the radiative cooling rate of inverse Compton
scattering. 
The main purpose of this paper is to present a model explaining the
bright/hard state.

In the conventional theory of accretion disks, the concept of phenomenological
$\alpha$-viscosity is introduced. In this framework, the $\varpi
\varphi$-component of the stress tensor, which appears in the angular
momentum equation and the viscous heating term, is assumed to be proportional to
the gas pressure, $t_{\varpi \varphi} = -\alpha_{\rm SS} p_{\rm
gas}$ (we ignore the radiation pressure in this paper because we focus
on optically thin disks). Here $\alpha_{\rm SS}$ is the viscosity
parameter introduced by \cite{shak73}. 
The magnetic field can be an origin of the $\alpha$-viscosity because
the Maxwell stress generated by the magneto-rotational instability (MRI)
efficiently transports angular momentum in accretion disks
\citep[e.g.,][]{balb91} and the dissipation of magnetic energy 
contributes the disk heating \citep[e.g.,][]{hiro06}.

Optically thin, hot accretion disks have been studied to explain hard
X-rays from BHCs. 
\cite{thor75} proposed that hard X-rays from Cyg X-1 are produced in an inner
optically thin hot disk. \cite{shib75} studied the structure and
stability of optically thin hot accretion disks. 
We have to consider two-temperature plasma in such
disks because the electron temperature is expected to become lower
than the ion temperature in such a low density, high temperature
region. The energy equations for ions and electrons are written in the
form
\begin{eqnarray}
 \rho_{\rm i} T_{\rm i} \frac{d S_{\rm i}}{d t}  = \left(1 - \delta_{\rm heat}
\right) q^{+} - q^{\rm ie} ~,
 \label{eq:org_ene_i}
\end{eqnarray}
\begin{eqnarray}
 \rho_{\rm e} T_{\rm e} \frac{d S_{\rm e}}{d t} = \delta_{\rm heat} q^{+} +
 q^{\rm ie} - q_{\rm rad}^{-} ~,
 \label{eq:org_ene_e}
\end{eqnarray}
where $q^{+}$ is the viscous heating rate, $q^{\rm ie}$ is the energy transfer
rate from ions to electrons via Coulomb collisions, $q_{\rm rad}^{-}$ is
the radiative cooling rate, and $\delta_{\rm heat}$ is the fraction of
heating to 
electrons. The left hand-sides represent the
heat advection terms for ions ($q_{\rm ad,i}$) and electrons ($q_{\rm
ad,e}$). We note that early works on optically
thin, hot, two-temperature accretion 
disks assumed that the viscous heating acts primarily on ions ($\delta_{\rm heat}
\ll 1$).

\cite{eard75} and
\citet[][hereafter SLE]{shap76} constructed a model for optically thin
two-temperature accretion disks. In the SLE solutions, the dissipated
energy is transferred from ions to electrons via Coulomb
collisions ($q^{+} \sim q^{\rm ie}$) and radiated away by electrons
($q^{\rm ie} \sim 
q_{\rm rad}^{-}$). Although the electron temperature is high
enough to explain the X-ray spectrum in the low/hard state, the SLE solutions
are thermally unstable in the framework of the $\alpha$-prescription of
viscosity.

\cite{ichi77} pointed out the importance of heat advection in hot,
magnetized accretion flows, and obtained steady solutions
of optically thin disks. Such geometrically thick, optically thin,
advection-dominated accretion flows (ADAFs) or radiatively inefficient
accretion flows (RIAFs) have been studied
extensively by \citet[][\citeyear{nara95}]{nara94} and
\cite{abra95}. In the ADAF/RIAF solutions, a substantial fraction of
the dissipated energy is stored in the gas as entropy and advected into
the central object ($q^{+} \sim q_{\rm ad,i}$). Only a small fraction
of the dissipated energy is transferred to electrons and radiated
away. The ADAF/RIAF solutions are thermally stable.
\cite{esin97} found that the maximum mass accretion rate of the ADAF/RIAF
solutions is ${\dot M}_{\rm c,A} \sim 1.3 \alpha^2 ~ {\dot M}_{\rm Edd}$
which corresponds to $L \sim 0.4 \alpha^2 L_{\rm Edd}$, where ${\dot
M}_{\rm Edd}$ is the Eddington mass accretion rate. 
\cite{esin98} showed that the electron temperature in the ADAF/RIAF
solutions weakly anti-correlates with the luminosity and decreases to $\sim
10^{9.5} {\rm K}$. These features are consistent
with the facts that the energy cutoff weakly anti-correlates with the
luminosity around $\sim 200 ~ {\rm keV}$ in the low/hard state, and that
these systems undergo a transition from the low/hard state to other
X-ray spectral states (i.e., the high/soft state during the dark/fast
transition, and the bright/hard state during the bright/slow transition)
about at this maximum luminosity of the ADAF/RIAF solutions. However,
the ADAF/RIAF solutions 
cannot explain the strong anti-correlation in the range of $L \gtrsim
0.1 ~ L_{\rm Edd}$ and $T_{\rm e} \lesssim 200 ~ {\rm keV}$ observed in the
bright/hard state.

The heat advection works as an effective cooling in the ADAF/RIAF solutions. 
\cite{yuan01,yuan03a} 
presented a luminous hot accretion flow (LHAF) in which the heat
advection for ions works as an effective heating. Above the maximum
mass accretion rate of the
ADAF/RIAF solutions, the heat advection overwhelms the viscous heating
and balances the energy transfer from ions to electrons ($q_{\rm ad,i}
\sim q^{\rm ie}$). 
The LHAFs are thermally unstable. However, \cite{yuan03b} concluded that
the thermal instability will have no effect on the dynamics of the LHAFs
because the accretion timescale is shorter than the timescale of
growth of the local perturbation at such high mass accretion rate. The
LHAF solutions also cannot explain the bright/hard state because the
electron temperature is high and roughly constant at $\sim 10^{9.5} ~ {\rm
K}$.

In these models mentioned above, magnetic fields are not considered
explicitly, and the ratio of the gas pressure to the magnetic pressure
is assumed to be constant (typically, $\beta = p_{\rm gas} / p_{\rm mag}
\sim 1$). \cite{shib90} suggested that an accretion disk evolves toward
two types of disks, a high-$\beta$ disk and a low-$\beta$ disk, by
carrying out two-dimensional magnetohydrodynamic (MHD) simulations of
the buoyant escape of the magnetic flux owing to the Parker instability
\citep{park66}. In the high-$\beta$ disk, the magnetic flux escapes from
the disk owing to the Parker instability and $\beta$ inside the disk is
maintained at a high value. Global three-dimensional MHD simulations of
optically thin, radiatively inefficient accretion disks also indicate
that the amplification of magnetic fields becomes saturated when $\beta
\sim 10$ in a quasi-steady state except in the plunging region very
close to the black hole
\citep[e.g.,][]{hawl00,mach00,hawl01,mach03,mach04}. On the other hand, once a
disk is dominated by the magnetic pressure, it can stay in the
low-$\beta$ state because the strong magnetic tension suppresses the
growth of the Parker instability.

\cite{mach06} demonstrated that an
optically thin, radiatively inefficient, hot, high-$\beta$
(ADAF/RIAF-like) disk undergoes transition to an optically thin, radiatively
efficient, cool, low-$\beta$ disk except in the
plunging region when the mass accretion rate exceeds
the threshold for the onset of a cooling instability. During this
transition, the magnetic flux $\langle B_{\varphi} \rangle H$ is almost
conserved at each radius because the cooling timescale is shorter 
than that of the buoyant escape of the magnetic flux, where $\langle B_{\varphi}
\rangle$ is the mean azimuthal magnetic field and $H$ is the half
thickness of the disk. 
In this way, the magnetic pressure becomes dominant and
supports the disk as the gas pressure decreases owing
to the cooling instability. Eventually, the disk stays in a quasi-steady, cool,
low-$\beta$ state. Because the MRI is not yet stabilized in this
quasi-steady state, the magnetic field still
remains turbulent and dominated by the azimuthal component. As a result,
the heating owing to the dissipation of the turbulent magnetic field
balances the radiative cooling.

\cite{joha08} performed local three-dimensional MHD simulations of
strongly magnetized, vertically stratified accretion disks in a
Keplerian potential. They showed that strongly magnetized state is
maintained near the equatorial plane because the buoyantly escaping
magnetic flux is replenished by stretching of a radial field.
The MRI feeds off both vertical and azimuthal fields and drives
turbulence. The Maxwell and Reynolds stresses generated by the
turbulence become significant. Therefore, they indicated that highly
magnetized disks are astrophysically viable.

We note that such low-$\beta$ disks are quite different from
magnetically dominated accretion flows \citep[MDAFs;][]{meie05} observed
in the plunging region of optically thin accretion disks in global MHD
and general relativistic MHD simulations  \citep[e.g.,][]{frag09}
in terms of their energy balance and configuration of magnetic
fields. Outside the plunging region, the magnetic field
becomes turbulent and dominated by the azimuthal component because the
growth timescale of the MRI is shorter than the inflow timescale. As a
result, the released gravitational energy is efficiently converted into the
thermal energy via the dissipation of the turbulent magnetic field.
On the other hand in the plunging 
region, the ratio of the timescales is reversed because the inflow
velocity increases with decreasing the radius in such disks. Therefore, magnetic
field lines are stretched out in the radial direction before turbulence
is generated by the MRI and is dissipated. 
As a result, a substantial fraction of the
gravitational energy is converted into the radial infall kinetic
energy. Since the heating owing to the dissipation of turbulent magnetic
fields becomes inefficient, the gas pressure becomes low, and the flow
becomes magnetically dominated. Although both of the low-$\beta$ disk
and the MDAF are cool and magnetic pressure dominant, they are
essentially different. We focus on the low-$\beta$ disk in this paper.

\cite{oda07} constructed a steady model of optically thin,
one-temperature accretion disks incorporating magnetic fields on the
basis of these results of three-dimensional MHD simulations of accretion
disks. \cite{oda09} extended it to
the optically thick regime. They assumed that the $\varpi
\varphi$-component of the stress tensor is
proportional to the total pressure. In order to complete the set of
basic equations, they specified the advection rate of the 
azimuthal magnetic flux instead of $\beta$. They found
a new thermally stable solution, a low-$\beta$ solution, which can
explain 
the results by \cite{mach06}. In the low-$\beta$ solutions, the magnetic
heating enhanced by the strong magnetic pressure balances the radiative
cooling. The disk temperature is lower than that in the ADAF/RIAF solutions
and strongly anti-correlates with the mass accretion rate. They also found that
the low-$\beta$ solutions exist above the maximum mass accretion rate of
the ADAF/RIAF solutions. Therefore, they concluded that the optically thin
low-$\beta$ disk can 
qualitatively explain the bright/hard state. However, they considered
one-temperature plasma and bremsstrahlung emission as a radiative
cooling mechanism. It is expected that the electron
temperature becomes lower than the ion temperature and that synchrotron
cooling and/or inverse Compton scattering become effective in such disks.

In this paper, we extend the model of optically thin, one-temperature
disk to that of optically thin, two-temperature disks. We consider
synchrotron emission and inverse Compton scattering as a cooling mechanism
in addition to bremsstrahlung emission. 
We obtained the thermal equilibrium curves and found
that the optically thin low-$\beta$ solutions can quantitatively explain
the bright/hard state. The basic equations are presented in Section
\ref{basic_eq}. In Section \ref{results}, we present the thermal
equilibrium curves. Section \ref{discussion} is devoted to a
discussion. We summarize the paper in Section \ref{summary}. 

\section{Models and Assumptions} \label{basic_eq}

  \subsection{Basic Equations}
We extended the basic equations for one-dimensional steady,
optically thin, two-temperature black hole accretion flows
\cite[e.g.,][]{kato08} incorporating magnetic fields. We adopt
cylindrical coordinates $(\varpi,\varphi,z)$. General relativistic effects are
simulated using the pseudo-Newtonian potential $\psi =
-GM/(r-r_{\rm s})$ \citep{pacz80}, where $G$ is the
gravitational constant, $M$ is the black hole mass (we assume $M
= 10 M_\sun$ in this paper), $r = (\varpi^2 + z^2)^{1/2}$, and $r_{\rm s} =
2GM/c^2 $ is the Schwarzschild radius. For simplicity, the gas is
assumed to consist of protons (ions) and electrons. The number density of
ions and electrons are equal by charge
neutrality, $n = n_{\rm i} = n_{\rm e}$. 

We start with the resistive MHD equations
\begin{eqnarray}
 \frac{\partial \rho}{\partial t} + \nabla \cdot \left( \rho
 \mbox{\boldmath{$v$}} \right) = 0 ~,
 \label{eq:vec_con}
\end{eqnarray}
\begin{eqnarray}
 \rho \left[ \frac{\partial \mbox{\boldmath{$v$}}}{\partial t} + \left(
 \mbox{\boldmath{$v$}} 
 \cdot \nabla \right) \mbox{\boldmath{$v$}}
   \right] = 
 - \rho \nabla \psi - \nabla p_{\rm gas} + \frac{\mbox{\boldmath{$j$}} \times
\mbox{\boldmath{$B$}}}{c} ~,
 \label{eq:vec_mom}
\end{eqnarray}
\begin{eqnarray}
 \frac{\partial \left( \rho_{\rm i} \epsilon_{\rm i} \right)}{\partial
  t} + \nabla \cdot \left[ \left(
 \rho_{\rm i} \epsilon_{\rm i} + p_{\rm i} \right) \mbox{\boldmath{$v$}}
 \right] - \left(
 \mbox{\boldmath{$v$}} \cdot \nabla \right) p_{\rm i} \nonumber \\
 = \left(1 -
 \delta_{\rm heat}
\right) q^{+} - q^{\rm ie} ~,
 \label{eq:vec_ene_i}
\end{eqnarray}
\begin{eqnarray}
 \frac{\partial \left( \rho_{\rm e} \epsilon_{\rm e} \right)}{\partial
  t} + \nabla \cdot  \left[ \left(
 \rho_{\rm e} \epsilon_{\rm e} + p_{\rm e} \right) \mbox{\boldmath{$v$}}
 \right] - \left(
 \mbox{\boldmath{$v$}} \cdot \nabla \right) p_{\rm e} \nonumber \\
 = \delta_{\rm
 heat} q^{+} +
 q^{\rm ie} - q_{\rm rad}^{-} ~,
 \label{eq:vec_ene_e}
\end{eqnarray}
\begin{eqnarray}
 \frac{\partial \mbox{\boldmath{$B$}}}{\partial t} = \nabla \times
  \left( \mbox{\boldmath{$v$}}
 \times \mbox{\boldmath{$B$}} - \frac{4 \pi}{c}\eta_{\rm m}
 \mbox{\boldmath{$j$}} \right) ~,
 \label{eq:vec_ind}
\end{eqnarray}
where $\rho = \rho_{\rm i} + \rho_{\rm e}$ is the density, $\rho_{\rm i}
= m_{\rm i} n$ and $\rho_{\rm e} = m_{\rm e} n$ are 
the ion and electron densities, $m_{\rm i}$ and $m_{\rm e}$ are the
ion and electron masses, $\mbox{\boldmath{$v$}}$ is the velocity,
$\mbox{\boldmath{$B$}}$
is the magnetic field, $\mbox{\boldmath{$j$}} = c \nabla \times
\mbox{\boldmath{$B$}} / 4 \pi$
is the current density, $p_{\rm gas} = p_{\rm i} + p_{\rm e} = n k
\left(T_{\rm i} + T_{\rm e} \right)$ is the gas pressure, $p_{\rm i}$
and $p_{\rm e}$ are the ion and electron gas pressure, $T_{\rm i}$ and
$T_{\rm e}$ are the ion and electron temperature, $k$ is the Boltzmann
constant, $\epsilon_{\rm i} = 
\left( p_{\rm i} / \rho_{\rm i} \right) / \left( \gamma_{\rm i} -1
\right)$ and $\epsilon_{\rm e} = 
\left(p_{\rm e} / \rho_{\rm e}\right) / \left(\gamma_{\rm e}
-1
\right)$ are the internal energy of ions and electrons. Here,
$\gamma_{\rm i} = 5/3$ and $\gamma_{\rm e} = \gamma_{\rm e} \left(T_{\rm
e}\right)$ are the specific heat ratio for ions and electrons.
In the energy equations for ions (\ref{eq:vec_ene_i}) and electrons
(\ref{eq:vec_ene_e}), $q^{+}$ is the heating rate, $q_{\rm rad}^{-}$ is the 
radiative cooling rate, and $q^{\rm ie}$ is the energy transfer rate from
ions to electrons via Coulomb collisions. Here,
$\delta_{\rm heat}$ represents the fraction of heating to electrons. In the
induction equation (\ref{eq:vec_ind}),
$\eta_{\rm m} \equiv c^2/4 \pi \sigma_{\rm c}$ is the magnetic diffusivity, where
$\sigma_{\rm c}$ is the electric conductivity.

 \subsubsection{Azimuthally Averaged Equations}
Three-dimensional global MHD and local radiation-MHD simulations of
black hole accretion
disks showed that magnetic fields inside the disk are turbulent and
dominated by the azimuthal component in a quasi-steady state
\cite[e.g.,][]{mach06, hiro06}. 
On the basis of results of the simulations, we decomposed the magnetic
fields into the
mean fields $\mbox{\boldmath{$\bar{B}$}} =
\left( 0, \bar{B_{\varphi}}, 0 \right)$ and 
fluctuating fields $\delta \mbox{\boldmath{$B$}} = \left( \delta B_{\varpi},
 \delta B_{\varphi}, \delta B_{z} \right)$ and also decomposed the
velocity into the mean velocity $\mbox{\boldmath{$\bar{v}$}} =
(v_{\varpi}, v_{\varphi}, v_{z})$ and the
fluctuating velocity $\delta \mbox{\boldmath{$v$}} = \left(\delta v_{\varpi},
 \delta v_{\varphi}, \delta v_{z} \right)$. We assume that the
fluctuating components vanish when azimuthally averaged, $\langle \delta
\mbox{\boldmath{$v$}} \rangle = \langle \delta \mbox{\boldmath{$B$}}
\rangle = 0$, and that the 
radial and vertical components of the magnetic fields are negligible
compared with that of the azimuthal component, $|\bar{B_{\varphi}}
 + \delta B_{\varphi}| \gg |\delta B_{\varpi}|$, $|\delta
B_{z}|$. Here $\langle ~~ \rangle$
denotes the azimuthal average.

Let us derive the azimuthally averaged equations. We assume that the disk is
 in a steady state and in hydrostatic balance in the vertical
direction. By azimuthally averaging equations (\ref{eq:vec_con}) -
(\ref{eq:vec_ind}) and ignoring the second order terms of $\delta
 \mbox{\boldmath{$v$}}$, $\delta B_{\varpi}$,
 and $\delta B_{z}$, we obtain
\begin{equation}
 \label{eq:con}
 \frac{1}{\varpi} \frac{\partial}{\partial \varpi} \left( \varpi \rho
 v_{\varpi} \right) +
 \frac{\partial}{\partial z} \left( \rho v_{z} \right) = 0 ~,
\end{equation}
\begin{equation}
 \label{eq:mom_pi}
 \rho v_{\varpi} \frac{\partial v_{\varpi}}{\partial \varpi} + \rho
 v_{z} \frac{\partial v_{\varpi}}{\partial z} - \frac{\rho
 v_{\varphi}^{2}}{\varpi} = - \rho \frac{\partial \psi}{\partial
 \varpi} - \frac{\partial p_{\rm tot}}{\partial \varpi}
- \frac{\langle B_{\varphi}^2 \rangle}{4 \pi \varpi } ~, 
\end{equation}
\begin{eqnarray}
 \label{eq:mom_phi}
 \rho v_{\varpi} \frac{\partial v_{\varphi}}{\partial \varpi} + \rho
 v_{z} \frac{\partial v_{\varphi}}{\partial z} + \frac{\rho
 v_{\varpi} v_{\varphi}} {\varpi} \nonumber \\
 = \frac{1}{{\varpi}^{2}}
 \frac{\partial}{\partial \varpi} \left[ {\varpi}^{2}
\frac{ \langle B_{\varpi} B_{\varphi} \rangle}{4\pi}
 \right] + \frac{\partial}{\partial z}
 \left( \frac{\langle B_{\varphi} B_{z} \rangle}{4 \pi}\right)~,
\end{eqnarray}
\begin{equation}
 \label{eq:mom_z}
 0 = - \frac{\partial \psi}{\partial z}
 - \frac{1}{\rho} \frac{\partial p_{\rm tot}}{\partial z}
 ~,
\end{equation}
\begin{eqnarray}
 \label{eq:ene_i}
 \frac{\partial}{\partial \varpi} \left[ \left( \rho_{\rm i} \epsilon_{\rm i} +
					   p_{\rm i}\right)
 v_{\varpi}\right]
 + \frac{v_{\varpi}}{\varpi} \left( \rho_{\rm i} \epsilon_{\rm i} +
 p_{\rm i} \right) +
 \frac{\partial}{\partial z} \left[
 \left( \rho_{\rm i} \epsilon_{\rm i} + p_{\rm i} \right) v_{z}\right]
 \nonumber \\ 
 - v_{\varpi} \frac{\partial}{\partial \varpi} p_{\rm i} - v_{z}
  \frac{\partial}{\partial z} p_{\rm i} = \left( 1 - \delta_{\rm heat} \right) q^{+} -
  q^{\rm ie} ~,
\end{eqnarray}
\begin{eqnarray}
 \label{eq:ene_e}
 \frac{\partial}{\partial \varpi} \left[ \left( \rho_{\rm e} \epsilon_{\rm e} +
					   p_{\rm e}\right)
 v_{\varpi}\right]
 + \frac{v_{\varpi}}{\varpi} \left( \rho_{\rm e} \epsilon_{\rm e} +
 p_{\rm e} \right) +
 \frac{\partial}{\partial z} \left[
 \left( \rho_{\rm e} \epsilon_{\rm e} + p_{\rm e} \right) v_{z}\right]
 \nonumber \\ 
 - v_{\varpi} \frac{\partial}{\partial \varpi} p_{\rm e} - v_{z}
  \frac{\partial}{\partial z} p_{\rm e} = \delta_{\rm heat} q^{+} + q^{\rm ie} -
  q^{-}_{\rm rad} ~,
\end{eqnarray}
\begin{eqnarray}
 \label{eq:ind_phi}
 0 = -\frac{\partial}{\partial z} \left[ v_{z} \langle B_{\varphi} \rangle
 \right] -\frac{\partial}{\partial \varpi} \left[ v_{\varpi} \langle
 B_{\varphi} \rangle \right]  \nonumber \\ 
 + \{ \nabla \times \langle \delta
  \mbox{\boldmath{$v$}} \times \delta \mbox{\boldmath{$B$}} \rangle
 \}_{\varphi} - \{ \eta_{\rm m} \nabla \times \left( \nabla \times
  \mbox{\boldmath{$\bar{B}$}} \right) \}_{\varphi} ~,
\end{eqnarray}
where $p_{\rm tot} = p_{\rm gas} + p_{\rm mag}$ is the total
pressure and $ p_{\rm mag} = \langle B_{\varphi}^{2} \rangle /8 \pi$ is the
azimuthally averaged magnetic pressure. The third and fourth terms on the
right-hand side of Equation (\ref{eq:ind_phi}) represent the dynamo term
and the magnetic
diffusion term which we approximate later on the basis of
the results of the
numerical simulations.

\subsubsection{Vertically Integrated, Azimuthally Averaged Equations}

We assume that the radial velocity $v_{\varpi}$, the specific angular momentum
$\ell = \varpi v_{\varphi}$, and $\beta \equiv p_{\rm gas} /p_{\rm mag}$
are independent of $z$, and that the disks are isothermal in
the vertical direction. 
Under these assumptions, the surface density $\Sigma$, the vertically integrated
total pressure $W_{\rm tot}$, and the half thickness of the disk $H$ are
defined as 
\begin{eqnarray}
 \label{eq:si}
 \Sigma \equiv \int_{-\infty}^{\infty}\rho dz =
 \int_{-\infty}^{\infty}\rho_0\exp\left(-\frac{1}{2}\frac{z^2}{H^2}\right)
 dz = \sqrt{2 \pi} \rho_0 H ~,
\end{eqnarray}
\begin{eqnarray}
 \label{eq:wtot}
 W_{\rm tot} &\equiv& \int_{-\infty}^{\infty} p_{\rm tot}dz  \nonumber \\ 
 &=& \int_{-\infty}^{\infty}p_{{\rm tot}
  0} \exp \left(-\frac{1}{2}\frac{z^2}{H^2} \right) dz = \sqrt{2 \pi}
  p_{{\rm tot} 0} H ~, 
\end{eqnarray}
\begin{eqnarray}
 \label{eq:h2}
 \Omega_{{\rm K} 0}^2 H^2 = \frac{W_{\rm tot} }{\Sigma} ~,
\end{eqnarray}
where $\Omega_{{\rm K}0}=(GM/\varpi)^{1/2}/(\varpi -r_{\rm s})$ is the Keplerian
angular velocity. Here the subscript $0$ refers to quantities in the
equatorial plane. Using the equation of state for the ideal gas, the
vertically integrated total pressure is expressed as 
\begin{eqnarray}
 \label{eq:eos}
 W_{\rm tot} = W_{\rm gas} + W_{\rm mag} = \frac{k T_{\rm i} + k T_{\rm
 e}}{m_{\rm i} + m_{\rm e}} \Sigma \left(1 + \beta^{-1} \right) ~. 
\end{eqnarray}

Now we integrate the other basic equations in the vertical direction.
We obtain
\begin{eqnarray}
 \label{eq:con_int}
 \dot{M} = -2\pi\varpi\Sigma v_\varpi ~ ,
\end{eqnarray}
\begin{eqnarray}
 \label{eq:mom_pi_int}
  v_{\varpi} \frac{\partial v_{\varpi}}{\partial \varpi} +
  \frac{1}{\Sigma} \frac{\partial W_{\rm tot}}{\partial \varpi}
  \nonumber \\ 
 =
  \frac{{\ell}^2 - \ell_{\rm K 0}^2}{\varpi^3} - \frac{W_{\rm tot}}{\Sigma}
  \frac{\partial \ln \Omega_{\rm K 0}}{\partial \varpi} -\frac{2
  \beta^{-1}}{1+\beta^{-1}} \frac{W_{\rm tot}}{\Sigma} \frac{1}{\varpi} ~,
\end{eqnarray}
\begin{eqnarray}
 \label{eq:mom_phi_int}
 \dot M(\ell - \ell_{\rm in})= -2\pi \varpi^2 \int_{-\infty}^{\infty} \frac{\langle
 B_{\varpi} B_{\varphi} \rangle }{4 \pi} dz ~,
\end{eqnarray}
\begin{eqnarray}
 \label{eq:ene_i_int}
 Q_{\rm ad,i}= \left( 1 - \delta_{\rm heat} \right) Q^{+} - Q^{\rm ie} ~,
\end{eqnarray}
\begin{eqnarray}
 \label{eq:ene_e_int}
 Q_{\rm ad,e} = \delta_{\rm heat} Q^{+} + Q^{\rm ie} - Q_{\rm rad}^{-} ~,
\end{eqnarray}
\begin{eqnarray}
 \label{eq:ind_int}
 \dot \Phi &\equiv& \int_{-\infty}^{\infty}v_{\varpi} \langle
 B_{\varphi} \rangle 
 dz  \nonumber \\ 
 &=& \int_{\varpi}^{\varpi_{\rm out}} \int_{-\infty}^{\infty}
 [ \{ \nabla \times \langle \delta
  \mbox{\boldmath{$v$}} \times \delta \mbox{\boldmath{$B$}} \rangle
 \}_{\varphi} \nonumber \\
 &-& \{ \eta_{\rm m} \nabla \times \left( \nabla \times
  \mbox{\boldmath{$\bar{B}$}} \right) \}_{\varphi} ] d\varpi dz +
\mbox{const.}
\end{eqnarray}
where $\dot M$ is the mass accretion rate, $\ell_{{\rm K} 0} = \varpi^2
\Omega_{{\rm K} 0}$ is the Keplerian angular momentum and $\ell_{\rm in}$ is the
specific angular momentum swallowed by the black hole. In the energy
equations, 
$Q_{\rm ad,i}$ and $Q_{\rm ad,e}$ are the vertically integrated heat
advection terms for ions and electrons, 
$Q^{+}$, $Q_{\rm rad}^{-}$, and $Q^{\rm ie}$ are
the vertically integrated heating rate, radiative cooling
rate, and energy transfer rate from ions to electrons via Coulomb
collisions. 
In Equation
(\ref{eq:ind_int}), $\dot \Phi$ is the radial advection rate of the
azimuthal magnetic flux (hereafter we call it the magnetic flux advection
rate).

In this paper, we assume that $\ell = \ell_{\rm K 0}$ instead of Equation
(\ref{eq:mom_pi_int}) because we focus on local thermal equilibrium
solutions.

\subsection{$\alpha$-Prescription of the Maxwell Stress Tensor}

Global MHD simulations of radiatively inefficient, 
accretion flows \citep[e.g.,][]{hawl01,mach06} showed that
the ratio of the azimuthally 
averaged Maxwell stress to the sum of the azimuthally averaged gas
pressure and magnetic pressure is
nearly constant ($\alpha_{\rm B} 
\equiv - \langle B_{\varpi} B_{\varphi} / 4 \pi \rangle / \langle
p_{\rm gas} + p_{\rm mag}
\rangle \sim 0.05 - 0.1$) except in the plunging region. 
Following the simulation results, we assume that the azimuthally
averaged $\varpi \varphi$-component of the Maxwell stress inside a disk is
proportional to the total (gas and magnetic) pressure
\begin{eqnarray}
 \label{eq:al}
\frac{\langle B_{\varpi} B_{\varphi} \rangle }{4 \pi} = - \alpha p_{\rm tot}~.
\end{eqnarray}
Integrating in the vertical direction, we obtain
\begin{eqnarray} 
 \label{eq:al_int}
\int_{-H}^{H} \frac{\langle B_{\varpi} B_{\varphi} \rangle }{4 \pi} dz= -
\alpha W_{\rm tot}~.
\end{eqnarray}
This is one of the key assumptions in this paper. When the magnetic
pressure is high,
the stress level can be high even though the gas pressure is low.

We can rewrite this relation
in terms of the kinematic viscosity, $\nu$, as
\begin{eqnarray}
 \label{eq2:nu}
  \nu = A_{\nu} \alpha
  \sqrt{{c_{{\rm s}0}}^2 + {c_{{\rm A}0}}^2} H
\end{eqnarray}
where
\begin{eqnarray}
 \label{eq2:anu}
  A_{\nu} \equiv - \left( \frac{\Omega}{\Omega_{{\rm K} 0}} \frac{d \ln
		    \Omega}{d \ln \varpi}\right)^{-1} ~,
\end{eqnarray}
$c_{{\rm s}0}=\sqrt{p_{{\rm gas}0}/ \rho_0}$ is the
sound speed, and $c_{{\rm A}0} = \sqrt{2 p_{{\rm mag}0}/\rho_0}$ is the
Alfv\'{e}n speed.

\subsection{Prescription of the Magnetic Flux Advection Rate}
We complete the set of basic equations by specifying the radial
distribution of the magnetic flux advection rate. If we 
perform the integration
in the second term of the induction equation (\ref{eq:ind_int}), we obtain
\begin{eqnarray}
 \label{eq:ind_int_ignore}
 \dot \Phi &\equiv& - v_{\varpi} B_0(\varpi) \sqrt{4 \pi} H \\
 &=&
 \left[\mbox{dynamo and diffusion terms}\right] + \mbox{const.} \nonumber
\end{eqnarray}
where
\begin{eqnarray}
 \label{eq:b0}
 B_0(\varpi) = 2^{5/4} \pi^{1/4}
 \left(\frac{k T_{\rm i} + k T_{\rm e}}{m_{\rm i} + m_{\rm
  e}} \right)^{1/2}\Sigma^{1/2}H^{-1/2}\beta^{-1/2}
\end{eqnarray}
is the mean azimuthal magnetic field in the equatorial
plane. 
According to the result of the global three-dimensional MHD 
simulation by \cite{mach06}, the magnetic flux advection rate at a
radius is roughly unchanged before and after the transition from the
ADAF/RIAF-like disk to the low-$\beta$ disk. Hence, we adopt the
magnetic flux advection rate as the parameter in order to complete the
set of the basic equations. The magnetic flux advection rate depends on
various mechanisms such as the escape of magnetic fluxes due to the
magnetic buoyancy, the regeneration of azimuthal magnetic fields by the
shear motion, the generation of magnetic turbulence through the MRI,
dissipation of magnetic fields due to the magnetic diffusivity, and magnetic
reconnection. If the sum of the dynamo term and the magnetic diffusion
term is zero in the whole region, the magnetic flux advection rate is
constant in the radial direction. The global three-dimensional MHD
simulation performed by \cite{mach06} indicated that the magnetic
advection rate increases with decreasing radius, specifically, $\dot
\Phi \propto \varpi^{-1}$, in the quasi steady state as a result of
magnetic dynamo and diffusivity processes. Because it is hard to
compute the dynamo term and the magnetic diffusion term explicitly from
the local quantities, we parameterize the dependence of $\dot \Phi$
on $\varpi$ by introducing a parameter, $\zeta$, as follows. 
\begin{eqnarray}
 \label{eq:phidot}
 \dot \Phi(\varpi; \zeta, \dot M) \equiv {\dot \Phi}_{\rm out}(\dot M) \left(
  \frac{\varpi}{\varpi_{\rm out}}\right)^{-\zeta} ~,
\end{eqnarray}
where ${\dot \Phi}_{\rm out}$ is the magnetic flux advection rate at the
outer boundary $\varpi = \varpi_{\rm out}$. When $\zeta = 0$, the
magnetic flux advection rate is constant in the radial direction. When
$\zeta > 0$, the magnetic flux advection rate increases with decreasing
radius \citep[See also Figure 1 in][]{oda09}. Here we determine the parameter ${\dot \Phi}_{\rm out}$ by imposing the
outer boundary condition, $T_{\rm i,out} = T_{\rm e,out} = T_{\rm
virial} = \left[ 
\left(m_{\rm i} + m_{\rm e}\right) c^2 /
3 k \right] \left(\varpi_{\rm out}
/ r_{\rm s} \right)^{-1}$ and $\beta_{\rm out} = 10$ at
$\varpi_{\rm out}=1000r_{\rm s}$.
This leads that ${\dot \Phi}_{\rm out} \propto {\dot M}^{1/2}$. In the local
model presented in this paper, the
value of $\zeta$ just means the amount of the magnetic flux advection
rate at a radius (e.g, $\dot \Phi (\varpi=5r_{\rm s}) = {\dot
\Phi}_{\rm out}$, $4.9 ~ {\dot\Phi}_{\rm out}$, $24 ~ {\dot\Phi}_{\rm
out}$ for $\zeta = 0$, $0.3$, $0.6$, respectively).

Equation (\ref{eq:phidot}) is the second key assumption in this
paper. Specifying the magnetic flux advection rate enables the magnetic
pressure to increase when the disk temperature decreases. By contrast, if we
specified the plasma $\beta$ at each radius instead of the magnetic flux
advection rate, the decrease in
the temperature results in a decrease in 
magnetic pressure. This is inconsistent with the results of
three-dimensional MHD simulations \citep[e.g.,][]{mach06}.

We address the similarity between the concepts of the mass accretion rate
and the magnetic flux advection rate
in order to facilitate understanding of the concept of
the magnetic flux advection rate. The right-hand side of
equation (\ref{eq:con_int}) represents the mass flux crossing at $\varpi$
per unit time and we have denoted it by $\dot M$. If there is
no mass loss or gain (e.g, due to inflows and/or outflows), $\dot M$ is
constant in the radial direction. Otherwise $\dot M$ is a function of
$\varpi$ (e.g., $\dot M 
\propto \varpi^s$, $s$ is a parameter). The right-hand side of
equation (\ref{eq:ind_int_ignore}) represents the azimuthal magnetic flux
crossing at $\varpi$ per unit time and we have denoted it by $\dot \Phi$. If
there is no loss or gain of azimuthal magnetic fields
(e.g, due to the dynamo and/or the magnetic diffusion), $\dot \Phi$ is
constant in the radial direction. Otherwise $\dot \Phi$ is a function of
$\varpi$ and we 
prescribed it as equation (\ref{eq:phidot}). That is, when $\zeta=0$,
the magnetic flux is conserved in the radial direction, and when
$\zeta>0$ (or $\zeta < 0$), the magnetic flux increases (or decreases)
with a decreasing radius \citep[see Figure 1(b) in][]{oda09}. We note
that the
azimuthal magnetic flux inside a disk can increase when the azimuthal
flux of opposite polarity buoyantly escapes from the disk
\citep[e.g.,][]{nish06}.

\subsection{Energy Equations}
\subsubsection{The Magnetic Heating Rate}
In the conventional theory, the viscous heating was
expressed as $q^{+}_{\rm vis} = t_{\varpi \varphi} \varpi \left( d \Omega / d
\varpi \right)$ and assumed to heat primarily ions ($\delta_{\rm heat} \sim
m_{\rm e}/m_{\rm i} \sim 10^{-3}$), where $t_{\varpi \varphi}$
is the $\varpi \varphi$-component of the total stress and $\Omega$ is
the angular velocity. 

The results of three-dimensional MHD simulations indicate that the dissipation
of the magnetic energy dominates the total dissipative heating rate throughout
a disk and is expressed as $q^{+} \sim \langle B_{\varpi} B_{\varphi}
/ 4 \pi \rangle \varpi \left( d \Omega / d \varpi \right)$
\citep[e.g.,][]{hiro06,mach06,krol07}. Hereafter, we refer to it as the magnetic
heating rate. Following these simulation results, we employ magnetic heating as
the heating mechanism inside a disk, and set the vertically integrated
heating rate as follows:

\begin{eqnarray}
 \label{eq:qmag}
 Q^{+} = \int^{\infty}_{-\infty} \left[ \frac{\langle B_{\varpi}B_{\varphi}
 \rangle}{4\pi} \varpi
 \frac{d \Omega}{d \varpi} \right] dz = - \alpha W_{\rm tot} \varpi
 \frac{d \Omega}{d \varpi} ~ ,
\end{eqnarray}
where we have used equation (\ref{eq:al}). 
We note that if the
magnetic pressure is high, the heating rate can also be
high even when the gas pressure is low. The eventual expression of the
heating term is not at all unusual
\citep[e.g.,][]{naka97,manm97,yuan01} except that we consider much lower
values of $\beta$. We will
discuss in Section {\ref{discussion_main}} a lower limit of
$\beta$ below which the MRI is stabilized \citep[e.g.,][]{pess05}
so that this expression is no longer valid.

We note that the magnetic heating does not always heat primarily
ions. \cite{yuan03b} suggested that $\delta_{\rm heat} \sim 0.1 - 0.5$
is required
to fit the spectrum of Sgr $\rm A^{*}$ with RIAF models. \cite{shar07}
carried out local shearing box simulations of the nonlinear evolution of
the MRI in a collisionless plasma considering pressure anisotropy and
showed that $\delta_{\rm heat} \sim (1+3\sqrt{T_{\rm i}/T_{\rm
e}})^{-1}$ \citep[note that the definition of $\delta_{\rm heat} =
q_{\rm e}^{+}/q^{+}$ is different from
that of $\delta_{\rm heat} = q_{\rm e}^{+}/q_{\rm i}^{+}$ in][]{shar07}. 
The exact value we choose for this parameter is not so
important, in particular, for the low-$\beta$ solutions. 
We find that the low-$\beta$ solutions presented in this paper practically
unchanged for any value of $\delta_{\rm heat} \gtrsim 0.1$. Therefore, we adopt
$\delta_{\rm heat} = (1+3\sqrt{T_{\rm i}/T_{\rm e}})^{-1}$ as a fiducial value.

\subsubsection{The Energy Transfer Rate from Ions to Electrons by
   Coulomb Collisions}
If the ion temperature is higher than the electron temperature, Coulomb
collisions transfer energy from ions to electrons. The
energy transfer 
rate from ions to electrons per unit volume via Coulomb collisions is
given by \cite{step83}
\begin{eqnarray}
 q^{\rm ie} = -\frac{3}{2} \frac{m_{\rm e}}{m_{\rm i}} n^2 \sigma_T c  \left(\ln
  \Lambda\right) \frac{k T_{\rm e} - k  T_{\rm i}} {K_2(1/ \theta_{\rm e})
  K_2(1/ \theta_{\rm i})} \nonumber \\
  \left[ \frac{2(\theta_{\rm e} + \theta_{\rm i})^2
	      +1}{\theta_{\rm e} + \theta_{\rm i}} K_1 
   \left( \frac{\theta_{\rm e} + \theta_{\rm i}}{\theta_{\rm e}
    \theta_{\rm i}} \right) + 2 K_0 
   \left( \frac{\theta_{\rm e} + \theta_{\rm i}}{\theta_{\rm e}
    \theta_{\rm i}} \right)\right] ,
\end{eqnarray}
where $\sigma_{\rm T}$ is the Thomson scattering cross section and $\ln
\Lambda$ is the Coulomb logarithm (roughly
$\ln \Lambda \sim 20$). $K_n$ are modified 
Bessel function of the second kind of the order $n$, respectively.
The dimensionless electron and ion temperatures are defined by
\begin{eqnarray}
 \theta_{\rm e} = \frac{k T_{\rm e}}{m_{\rm e} c^2} ~,~ \theta_{\rm i} =
  \frac{k T_{\rm i}}{m_{\rm i} c^2} . 
\end{eqnarray}

For technical reason, we use the following formula which uses no special
functions, and is accurate to within a factor of 2 when $\theta_{\rm i}
< 0.2$ \citep{derm91}
\begin{eqnarray}
 q^{\rm ie} =  -\frac{3}{2} \frac{m_{\rm e}}{m_{\rm i}} n^2
  \sigma_{\rm T} c  \ln 
  \Lambda (k T_{\rm e} - k  T_{\rm i}) \frac{(2 \pi)^{1/2} + (\theta_{\rm e} +
  \theta_{\rm i})^{1/2}}{(\theta_{\rm e} + \theta_{\rm i})} .
\end{eqnarray}
Integrating in the vertical direction, we obtain
\begin{eqnarray}
 Q^{\rm ie} & = &  -\frac{3}{2} \frac{m_{\rm e}}{m_{\rm i}}
  \frac{\Sigma^2}{2 (m_{\rm i}+m_{\rm e})^2 \sqrt{\pi} H} \sigma_{\rm T}
  c  \ln  \Lambda  \\
 \nonumber \\
& \times & \left\{
\begin{array}{cc}
\displaystyle \frac{k T_{\rm e} - k  T_{\rm i}} {K_2(1/ \theta_{\rm e})
  K_2(1/ \theta_{\rm i})}  \times \nonumber \\
 \left[ \frac{2(\theta_{\rm e} + \theta_{\rm i})^2 +1}{\theta_{\rm e} +
  \theta_{\rm i}} K_1 
   \left( \frac{\theta_{\rm e} + \theta_{\rm i}}{\theta_{\rm e}
  \theta_{\rm i}} \right) + 2 K_0 
 \left( \frac{\theta_{\rm e} + \theta_{\rm i}}{\theta_{\rm e} \theta_{\rm i}}
  \right)\right] & (\theta_{\rm i} > 0.2)\\ \\
\displaystyle   (k T_{\rm e} - k  T_{\rm i}) \frac{(2 \pi)^{1/2} +
  (\theta_{\rm e} + 
  \theta_{\rm i})^{1/2}}{(\theta_{\rm e} + \theta_{\rm i})} &
  (\theta_{\rm i} < 0.2) \nonumber
\end{array}\right.  ~ .
\end{eqnarray}

We note that $\theta_{\rm i} < 0.2$ in almost all solutions presented in
this paper.

\subsubsection{Radiative Cooling Rate}
We assume that the radiative cooling occurs through electrons and
consider bremsstrahlung, synchrotron, and Compton cooling by
bremsstrahlung and synchrotron photons as cooling processes. 
The vertically integrated radiative cooling rate is expressed as
\begin{eqnarray}
 Q_{\rm rad}^{-} = Q_{\rm br}^{-} + Q_{\rm sy}^{-} + Q_{\rm br,C}^{-} +
  Q_{\rm sy,C}^{-} ~.
\end{eqnarray}

Following \cite{nara95} \citep[see also][]{sven82,step83},
bremsstrahlung cooling rate per unit volume is 
\begin{eqnarray}
 q_{\rm br}^{-} = q_{\rm br,ei}^{-} + q_{\rm br,ee}^{-} = n^2
  \sigma_{\rm T} c \alpha_{\rm f} m_{\rm e} c^2 \left[F_{\rm
  ei}(\theta_{\rm e}) + F_{\rm
  ee}(\theta_{\rm e}) \right]~,
\end{eqnarray}
where the subscripts $\rm ei$ and $\rm ee$ denote the electron-ion and
electron-electron bremsstrahlung cooling rates, 
$\alpha_{\rm f}$ is fine-structure constant, the function
$F_{\rm ei}(\theta_{\rm e})$ and $F_{\rm ee}(\theta_{\rm e})$ have the
approximate form
\begin{eqnarray}
 F_{\rm ei}(\theta_{\rm e}) = 
\left\{
\begin{array}{cc}
  \displaystyle \frac{9 \theta_{\rm e}}{2 \pi} \left[ \ln ( 2 \eta
  \theta_{\rm e} + 
  0.48) + \frac{3}{2}\right] & (\theta_{\rm e} > 1) \\ \\
  \displaystyle 4 \left( \frac{2 \theta_{\rm e}}{\pi^3}\right)^{1/2} \left[ 1
  + 1.781 \theta_{\rm e}^{1.34}\right] & (\theta_{\rm e} < 1)
\end{array} \right. ~,
\end{eqnarray}
\begin{eqnarray}
 F_{\rm ee}(\theta_{\rm e}) = 
\left\{
\begin{array}{cc}
  \displaystyle \frac{9 \theta_{\rm e}}{\pi} ( \ln ( 2 \eta \theta_{\rm e}) +
  1.28) & (\theta_{\rm e} > 1) \\ \\ 
  \displaystyle \frac{5}{6 \pi^{3/2}} (44 - 3 \pi^2 )\theta_{\rm
   e}^{3/2} \times \nonumber \\
 ( 1
  + 1.1 \theta_{\rm e} + \theta_{\rm e}^2 - 1.25 \theta_{\rm e}^{5/2}) &
  (\theta_{\rm e} < 1)
\end{array}\right.  ~,
\end{eqnarray}
$\eta_{\rm E} = \exp(- \gamma_{\rm E}) $ and $\gamma_{\rm E} \approx 0.5772$ is
Euler's number. 
Integrating in the vertical direction, we obtain
\begin{eqnarray}
 Q_{\rm br}^{-} = \sigma_{\rm T} c \alpha_{\rm f} m_{\rm e} c^2 
  \frac{\Sigma^2}{\left(m_{\rm i} + m_{\rm e}\right)^2 
  \sqrt{\pi} H} \times \nonumber \\ ~
  [F_{\rm ei}(\theta_{\rm e}) + F_{\rm ee}(\theta_{\rm e})] ~.
\end{eqnarray}

The synchrotron emissivity of a
relativistic Maxwellian distribution of electrons in the optically thin
limit is given by \cite{pach70}:
\begin{eqnarray}
 \epsilon_{\rm sy} d \nu = 
 \frac{2 e^2 n_{\rm e}}{\sqrt{3} c} \frac{2 \pi \nu}{K_2 (1/
  \theta_{\rm e})}  I^{\prime}\left( x_{\rm M}\right) d \nu ~, 
\end{eqnarray}
where 
\begin{eqnarray}
 x_{\rm M} = \frac{2 \nu}{3 \nu_{\rm b} \theta_{\rm e}^2}, ~ \nu_{\rm b}
  = \frac{e B_0}{2 \pi m_{\rm e} c} ~ ,
\end{eqnarray}
with the fitting function $I^{\prime}\left( x_{\rm M}\right)$ given by
\cite{maha96} 
\begin{eqnarray}
 I^{\prime}\left( x_{\rm M}\right) =  \frac{4.0505}{x_{\rm M}^{1/6}}
 \left( 1 +\frac{0.40}{x_{\rm M}^{1/4}}
  +\frac{0.5316}{x_{\rm M}^{1/2}} \right) \times \nonumber \\
 \exp(-1.8899 x_{\rm M}^{1/3}) ~.
\end{eqnarray}
Integrating in the vertical direction, we obtain
\begin{eqnarray} 
 \label{eq4:e_sy}
 E_{\rm sy} d \nu =   \frac{2 e^2 }{\sqrt{3} c} \frac{\Sigma}{m_{\rm i}
  + m_{\rm e}} \frac{2 \pi
  \nu}{K_2 (1/ \theta_{\rm e})}  I^{\prime}\left( x_{\rm M}\right) d \nu ~.
\end{eqnarray}
We assumed that the emission below a critical frequency, $\nu_{\rm c}$,
is completely self-absorbed so that the emissivity can be approximated by the
blackbody emission from the surface of the disk.
Following \cite{esin96}, we estimate $\nu_{\rm c}$ as the
frequency at which the synchrotron emission from the region $\varpi \sim
\varpi + \Delta \varpi$ is equal to the blackbody emission (in the
Rayleigh-Jeans limit) from the upper and lower surfaces of the
region. 
This condition gives the equation
\begin{eqnarray}
 \label{eq:nuc}
 (2 \pi \varpi \Delta \varpi) E_{\rm sy} d \nu = 2 (2 \pi \varpi \Delta
  \varpi) 2 \pi \frac{\nu_{\rm c}^2}{c^2} k T_{\rm e} d \nu ~ .
\end{eqnarray}
We obtain $\nu_{\rm c}$ by solving this equation numerically. 
Integrating over frequency, we obtain 
\begin{eqnarray}
 Q_{\rm sy}^{-} 
 &=& \displaystyle 2 \int_{0}^{\nu_{\rm c}}  2 \pi
  \frac{\nu_{\rm c}^2}{c^2} k T_{\rm e} d \nu + \int_{\nu_{\rm
  c}}^{\infty} E_{\rm sy} d \nu 
  \nonumber \\
 &=& \frac{4 \pi \nu_{\rm c}^3 k T_{\rm e}}{3 c^2} + 
\frac{2 e^2 }{\sqrt{3} c} 
  \frac{\Sigma}{\left(m_{\rm i} + m_{\rm e}\right)} \frac{1}{K_2 (1/ \theta_{\rm e}) a_1^{1/6}} \nonumber \\ 
 &\times& \left[ \frac{1}{a_4^{11/2}} \Gamma \left(\frac{11}{2},
		       a_4 \nu_{\rm c}^{1/3}\right) +
 \frac{a_2}{a_4^{19/4}} \Gamma \left(\frac{19}{4}, a_4 \nu_{\rm
				c}^{1/3}\right) \right. \nonumber
 \\ 
 &+&  \frac{a_3}{a_4^4}(a_4^3 \nu_{\rm c} + 3 a_4^2
		       \nu_{\rm c}^{2/3} \nonumber \\ 
 &+& \left. 6 a_4 \nu_{\rm c}^{1/3} + 6)
		       \exp(-a_4 \nu_{\rm c}^{1/3}) \right],  
\end{eqnarray}
where the parameters $a_1$, $a_2$, $a_3$, and $a_4$ are defined as
\begin{eqnarray}
 a_1 = \frac{2}{3 \nu_{\rm b} \theta_{\rm e}^2} ,~ a_2 =
  \frac{0.4}{a_1^{1/4}} ,~
  a_3 = \frac{0.5316}{a_1^{1/2}} ,~ a_4 = 1.8899 a_1^{1/3} ,
\end{eqnarray}
and $\Gamma (a,x)$ is the incomplete gamma function
\begin{eqnarray}
 \Gamma (a,x) = \int_{x}^{\infty} t^{a-1} \exp(-t) dt.
\end{eqnarray}

We adopted the prescription for the Compton energy enhancement factor
$\eta$ described by \cite{derm91}, which is defined to be the 
average change in energy of a photon between injection and escape:
\begin{eqnarray}
 \eta & = & 1 + \frac{P(A-1)}{(1 - PA)} \left[1- \left( \frac{x}{3
						  \theta_{\rm e}} 
					     \right)^{-1 -\ln P / \ln
 A}\right] \nonumber \\
 & \equiv & 1 + \eta_1 -\eta_2 \left( \frac{x}{\theta_{\rm
				e}}\right)^{\eta_3} ~,
\end{eqnarray}
where
\begin{eqnarray}
 x & = & \frac{ h \nu}{m_{\rm e} c^2}~ , ~~~ \tau_{\rm es} =
  \frac{\kappa_{\rm es} 
  \Sigma}{2}~, \nonumber \\ 
 P & = & 1 - \exp (-\tau_{\rm es})~ , ~~~ A = 1 + 4 \theta_{\rm e} + 16
  \theta_{\rm e}^2 ~, 
  \nonumber \\ 
 \eta_1 & = & \frac{P (A-1)}{1-PA}~ , ~~~ \eta_2 = 3^{-\eta_3} \eta_1 ,
  \nonumber \\
   \eta_3 & = &-1 - \ln P / \ln A ~.
\end{eqnarray}
Here, $P$ is the probability that an
escaping photon is scattered, $A$ is the mean amplification factor in
the energy of a scattered photon when the 
scattering electrons have a Maxwellian velocity distribution of
temperature $\theta_{\rm e}$. 
Following \cite{nara95}, the vertically integrated Compton cooling by
bremsstrahlung and synchrotron photons are given respectively by
\begin{eqnarray}
 Q_{\rm br,C}^{-} = 3 \eta_1 Q_{\rm br}^{-} \times \nonumber \\
 \left\{ \left( \frac{1}{3} -
						     \frac{x_c}{3 
				    \theta_{\rm e}}\right) - \frac{1}{\eta_3 +
 1}\left[ \left( \frac{1}{3}\right)^{\eta_3 + 1} - \left( \frac{x_c}{3
	   \theta_{\rm e}}\right)^{\eta_3 + 1} \right]\right\} ~,
\end{eqnarray}
\begin{eqnarray}
 Q_{\rm sy,C}^{-} = Q_{\rm sy}^{-} \left[ \eta_1 -
		   \eta_2\left(\frac{x_c}{\theta_{\rm
			  e}}\right)^{\eta_3}\right] ~, 
\end{eqnarray}
where $x_c = h \nu_{\rm c} / m_{\rm e} c^2$.

\subsubsection{Heat Advection Term}
The vertically integrated heat advection terms for ions and electrons
are expressed as
\begin{eqnarray}
 \label{eq:qadvi}
 Q_{\rm ad,i} = \frac{\dot M}{2\pi \varpi^2} \frac{k T_{\rm i}}{m_{\rm i}
  + m_{\rm e}} \xi_{\rm i}~,
\end{eqnarray}
\begin{eqnarray}
 \label{eq:qadve}
 Q_{\rm ad,e} = \frac{\dot M}{2\pi \varpi^2} \frac{k T_{\rm e}}{m_{\rm i}
  + m_{\rm e}} \xi_{\rm e}~,
\end{eqnarray}
 where  
\begin{eqnarray}
 \label{eq:xi_i}
 \xi_{\rm i} = - 
 a_{\rm i} \frac{\partial \ln T_{\rm
  i}}{\partial \ln \varpi} + \frac{\partial \ln \Sigma}{\partial \ln
  \varpi} - \frac{\partial \ln H}{\partial \ln \varpi} ~,
\end{eqnarray}
\begin{eqnarray}
 \label{eq:xi_e}
 \xi_{\rm e} = - 
 a_{\rm e} (T_{\rm e}) \left[ 1 + \frac{\partial \ln a_{\rm e} (T_{\rm
		e})}{\partial \ln T_{\rm e}}\right] \frac{\partial \ln T_{\rm
  e}}{\partial \ln \varpi} \nonumber \\
 + \frac{\partial \ln \Sigma}{\partial \ln
  \varpi} - \frac{\partial \ln H}{\partial \ln \varpi}~, 
\end{eqnarray}
are dimensionless quantities of the order of unity (hereafter we call
them the entropy gradient parameter), $a_{\rm i} = 1/(1-\gamma_{\rm i}) =
3/2$, $a_{\rm e}(T_{\rm e}) = 1/(1-\gamma_{\rm e} (T_{\rm e}))$,
respectively. The positivity (negativity) of the entropy gradient
parameter means that the heat advection term works as an effective
cooling (heating).

The entropy gradient parameter for ions, $\xi_{\rm i}$, has a positive
value in ADAF/RIAF solutions, while it has a negative value in LHAF solutions. 
The global three-dimensional MHD simulations
\citep[e.g.,][]{mach06} and steady, vertically integrated,
one-dimensional transonic solutions \citep[e.g.,][]{naka96,naka97,oda07}
of optically thin black hole accretion flows indicate that $\xi_{\rm i}
\sim 1$. Following these results, we adopt $\xi_{\rm i} = 1$ as a
fiducial value in this paper. We note that a value
of $\xi_{\rm i}$ is not important in SLE and low-$\beta$ solutions
because the heat advection term is negligible compared to the other
terms. 

According to \cite{naka97}, $\xi_{\rm e}$ can have a positive and negative
value ($-0.5 \lesssim \xi_{\rm e} \lesssim 0.5$). However, they assumed
that no viscous heat goes into electrons ($\delta_{\rm heat} = 0$). 
In the energy equation for electrons, when $\delta_{\rm heat} \gtrsim 0.1$, the
dissipated magnetic energy term is typically greater than or comparable
to the heat advection term in the inner region of the disk.
Therefore the exact value we choose for this parameter is not so
important, in particular, for SLE and low-$\beta$ solutions in which
$\delta_{\rm heat} Q^{+}, Q^{\rm ie}, Q_{\rm rad}^{-} \gg Q_{\rm ad,e}$. We find
that the results presented in this paper are practically unchanged for
any value of $\xi_{\rm e}$ between $-0.5$ and
$0.5$ when $\delta_{\rm heat}
\gtrsim 0.1$. 
We show the results for $\xi_{\rm e} = 0.5$ in most part of this paper.

\section{Results} \label{results}
We solved the above basic equations at $\varpi = 5 r_{\rm s}$ for given
parameters $\dot M$, $\alpha$, $\zeta$, $\delta_{\rm heat}$, $\xi_{\rm i}$, and
$\xi_{\rm e}$. We obtained new thermal equilibrium solutions,
low-$\beta$ solutions, in addition to the ADAF/RIAF (for
positive $\xi_{\rm i}$), SLE, LHAF (for negative $\xi_{\rm i}$) solutions in the
optically thin regime. 

\subsection{Low-$\beta$ solutions} \label{results_main}
\begin{figure}
\epsscale{1.0}
\plotone{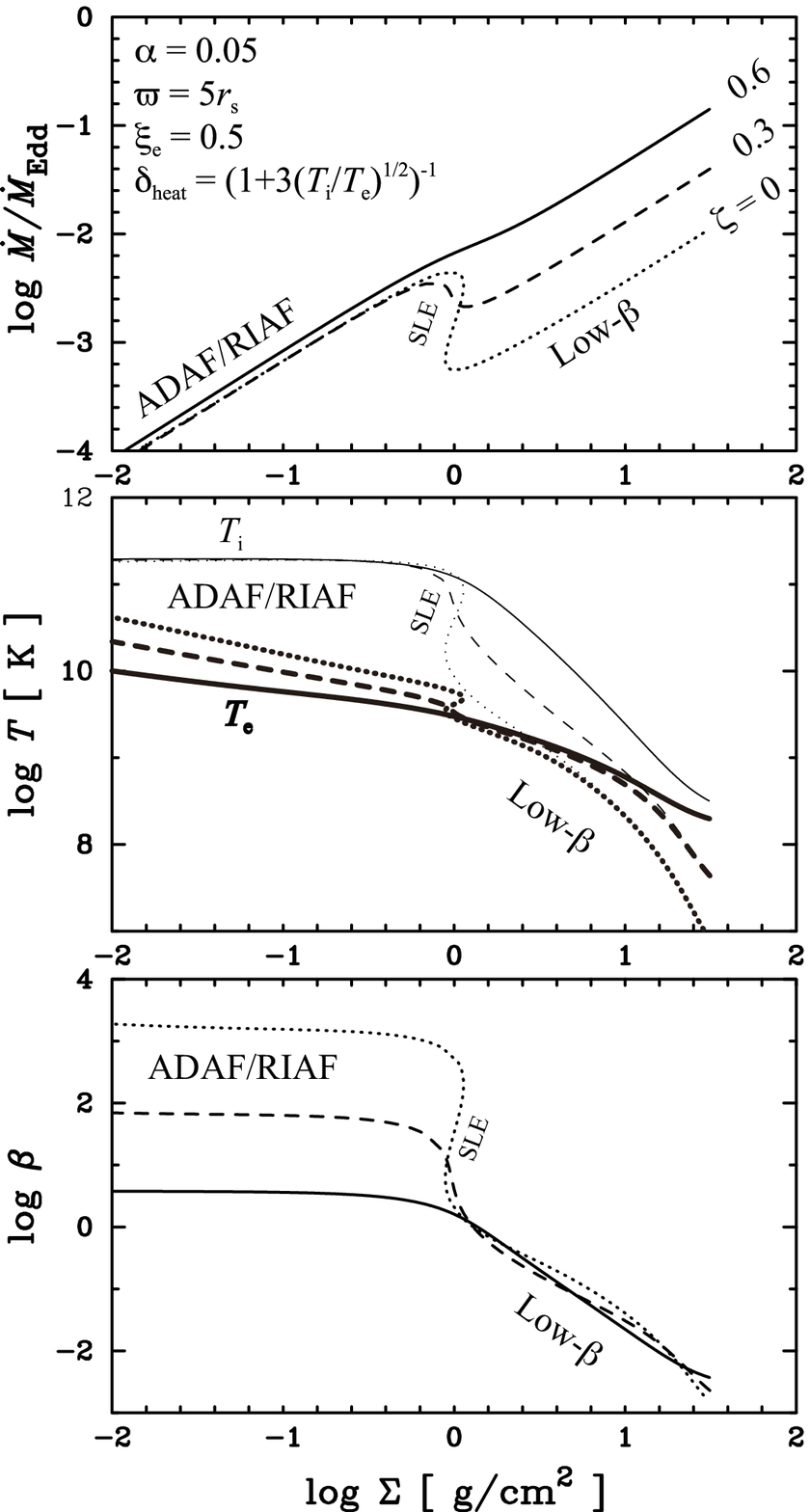}
\caption{Local thermal equilibrium curves of optically thin,
 two-temperature accretion disks at $\varpi = 5 r_{\rm s}$ in the $\Sigma$
vs. $\dot M / {\dot M}_{\rm Edd}$, $T_{\rm i}$(thin line), $T_{\rm
e}$(thick line), and $\beta$  plane. 
The disk parameters are $\alpha = 0.05$, $\xi_{\rm i} = 1$,
$\xi_{\rm e} = 0.5$, $\delta_{\rm heat} = (1+3\sqrt{T_{\rm i}/T_{\rm e}})^{-1}$,
$\zeta = 0.6$ (solid), $0.3$ (dashed), and $0$ (dotted). 
${\dot M}_{\rm Edd} = L_{\rm Edd} / \eta_{\rm e} c^2
= 4 \pi GM / \left( \eta_{\rm e} \kappa_{\rm es} c \right)$ is the
Eddington mass accretion rate and $\eta_{\rm e} = 0.1$ is the energy
conversion efficiency. The equilibrium curves consist of the ADAF/RIAF,
 SLE, and low-$\beta$ branches. \label{al05ds3xp05sifoo}}
\end{figure}

 Figure \ref{al05ds3xp05sifoo} shows the sequences of each thermal equilibrium
 solution in the $\Sigma$ versus $\dot M / {\dot M}_{\rm Edd}$, $T_{\rm
 i}$(thin line), $T_{\rm e}$(thick line), and $\beta$  plane.
The disk parameters we adopted are $\alpha = 0.05$, $\xi_{\rm i} = 1$,
$\xi_{\rm e} = 0.5$, $\delta_{\rm heat} = (1+3\sqrt{T_{\rm i}/T_{\rm e}})^{-1}$,
$\zeta = 0.6$ (solid), $0.3$ (dashed), and $0$ (dotted),
respectively. 
Here ${\dot M}_{\rm Edd} = L_{\rm Edd} / \eta_{\rm e} c^2
= 4 \pi GM / \left( \eta_{\rm e} \kappa_{\rm es} c \right)$ is the
Eddington mass accretion rate, $\eta_{\rm e} = 0.1$ is the energy
conversion efficiency, and $\kappa_{\rm es} = 0.40~ {\rm cm}^{2}~{\rm
g}^{-1}$ is the electron scattering opacity. 
We obtain three types of solutions, ADAF/RIAF (for $\Sigma \lesssim 1
~ {\rm g} ~ {\rm cm}^{-2}$ at this radius), SLE (for $\Sigma \sim 1 ~ {\rm g}
~ {\rm cm}^{-2}$), and low-$\beta$ solutions (for $\Sigma \gtrsim 1
~ {\rm g}~ {\rm cm}^{-2}$). 
We find that the low-$\beta$ solutions
exist above the maximum mass accretion rate of the ADAF/RIAF solutions,
${\dot M}_{\rm c, A} \sim 0.003 
{\dot M}_{\rm Edd}$. 
This indicates that the disk initially staying
in the ADAF/RIAF state undergoes transition to the low-$\beta$ state
when the mass accretion rate exceeds ${\dot M}_{\rm c, A}$. Furthermore,
the electron temperature in the low-$\beta$ solutions is lower ($T_{\rm
e} \sim 10^{8}-10^{9.5} {\rm K}$) than that in the ADAF/RIAF solutions.

\begin{figure}
\epsscale{1.0}
\plotone{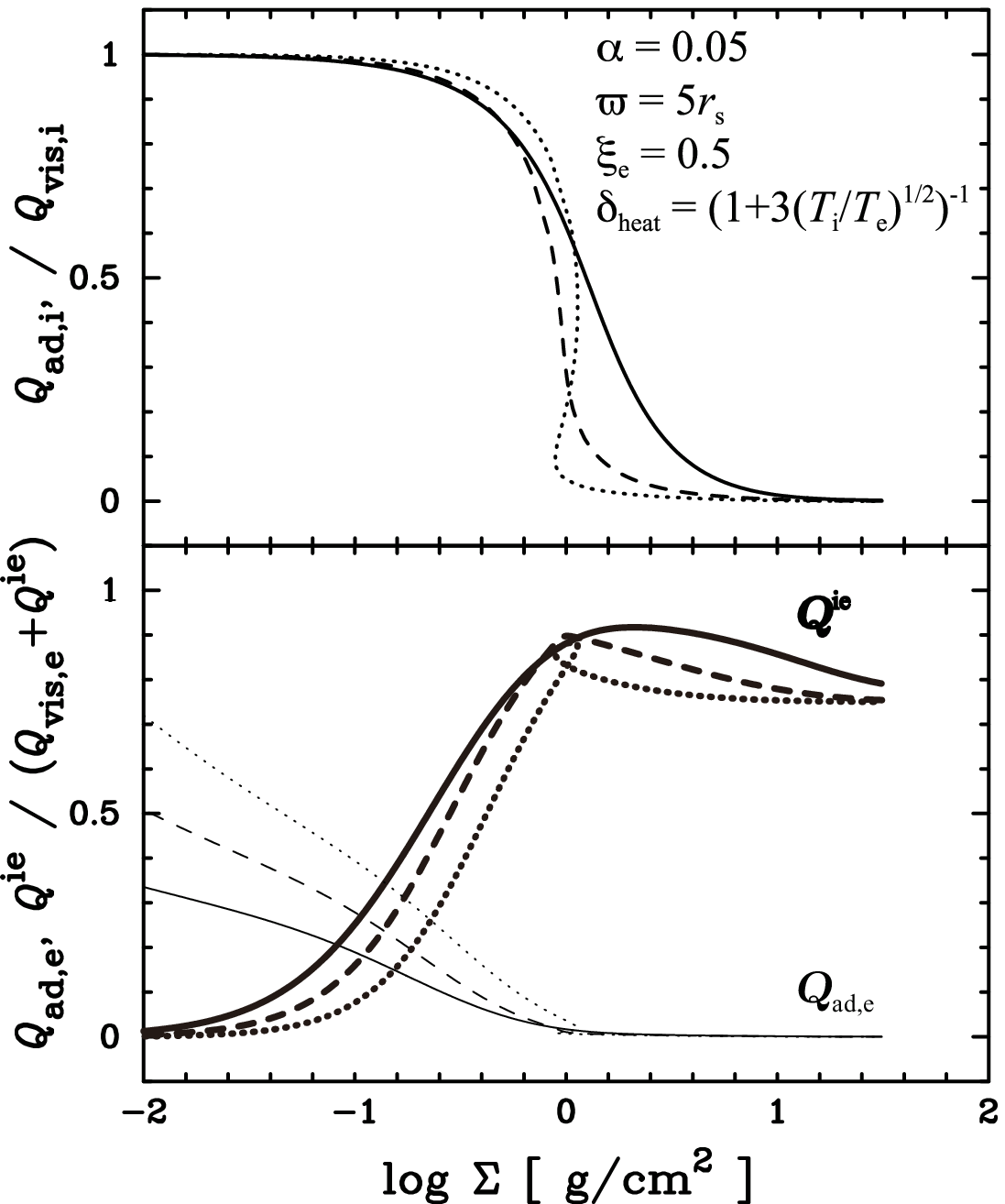}
\caption{Energy balance for ions and electrons for the same
 parameters as Figure \ref{al05ds3xp05sifoo} (solid: $\zeta = 0.6$,
 dashed: $\zeta = 0.3$,  dotted: $\zeta = 0$).  
Upper: the ratio of the heat advection to the magnetic
heating for ions, $Q_{\rm ad,i}/(1-\delta_{\rm heat})Q^{+}$ \cite[this parameter is
referred to as the advection factor $f$,
e.g.,][]{nara94,nara95,abra95,yuan01}. 
Lower: the ratio of the heat advection to the total heating
 (thin), $Q_{\rm ad,e}/(\delta_{\rm heat} Q^{+} + Q^{\rm ie})$, and
 the fraction of the energy transfer via Coulomb collisions to the total
 heating (thick), $Q^{\rm ie}/(\delta_{\rm heat} Q^{+} + Q^{\rm ie})$, for
 electrons. \label{al05ds3xp05sieie}}
\end{figure}

The energy balance for ions and electrons is illustrated in figure
\ref{al05ds3xp05sieie}. 
The upper panel shows the ratio of the heat advection to the magnetic
heating for ions, $Q_{\rm ad,i}/(1-\delta_{\rm heat})Q^{+}$ \cite[this quantity is
referred to as the advection factor $f$,
e.g.,][]{nara94,nara95,abra95,yuan01}. 
The lower panel shows the ratio of the heat advection to the total heating for
electrons (thin line) and the fraction of the energy transfer via
Coulomb collisions to the total heating (thick line). 
The electrons receive the total amount of the
magnetic heating in the low-$\beta$ solutions as well as in the SLE
solutions even
though we introduced the parameter $\delta_{\rm heat}$ which 
represents the fraction of heating to electrons. 
The fraction $(1-\delta_{\rm heat})$ of the magnetic heating goes into ions and the
fraction $\delta_{\rm heat}$ of the magnetic heating goes 
into electrons. However, almost all the magnetic heating going into ions is
transferred to electrons via Coulomb collisions in the low-$\beta$
solutions ($(1-\delta_{\rm heat})Q^{+} \sim Q^{\rm ie}$). Eventually, the total
amount of the magnetic heating goes into electrons. This means that the
parameter $\delta_{\rm heat}$ does not appear practically in the energy balance for
electrons. The radiative cooling overwhelms the heat advection in the
low-$\beta$ solutions. Therefore, the magnetic heating balances the
radiative cooling in the low-$\beta$ solutions ($Q^{+} \sim Q_{\rm
rad}^{-}$).

The energy balance is essentially the same in both the SLE solutions and the 
low-$\beta$ solutions ($Q^{+} \sim Q_{\rm rad}^{-}$). The difference is
which pressure dominates the magnetic heating. 
The magnetic heating, which is proportional to the total pressure in our
model, is dominated by the gas pressure in the SLE solutions and the
magnetic pressure in the low-$\beta$ solutions.

\begin{figure}
\epsscale{1.0}
\plotone{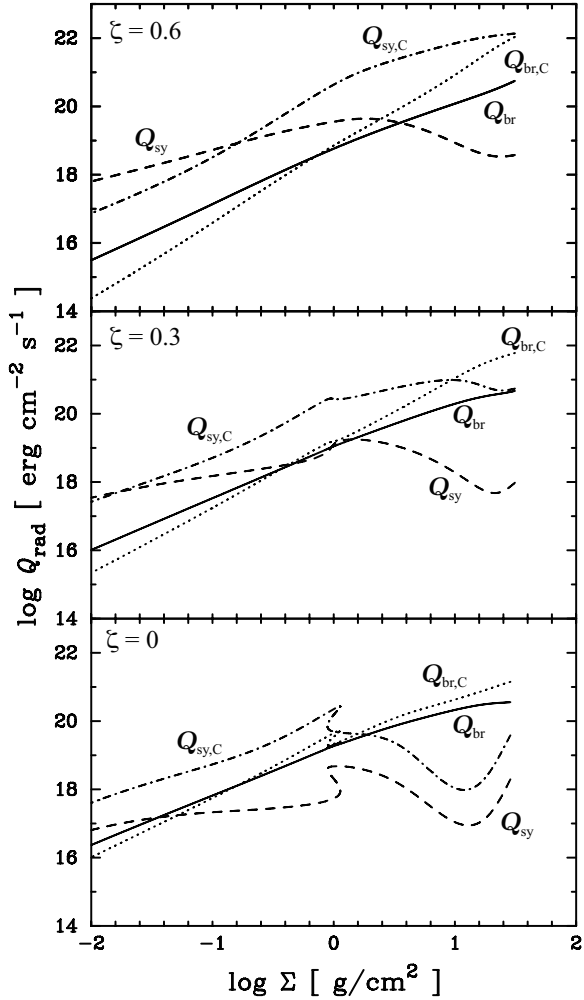}
\caption{Vertically integrated radiative cooling rates for $\zeta = 0$ (bottom),
$0.3$ (middle), and $0.6$ (top). 
Solid: bremsstrahlung ($Q_{\rm
 br}$), dotted: bremsstrahlung-Compton ($Q_{\rm br,C}$), dashed:
 synchrotron ($Q_{\rm sy}$), and dash-dotted: synchrotron-Compton
 ($Q_{\rm sy,C}$). \label{al05ds3xp05sirr}}
\end{figure}

Next we describe the main cooling mechanism in each solution. 
Figure \ref{al05ds3xp05sirr} shows 
the vertically integrated bremsstrahlung (solid),
bremsstrahlung-Compton (dotted), synchrotron (dashed), and
synchrotron-Compton (dash-dotted) cooling rate for $\zeta = 0$ (bottom),
$0.3$ (middle), and $0.6$ (top). 
When $\zeta = 0$, 
the synchrotron-Compton cooling is dominant in the low-$\beta$ solutions
 for lower mass accretion rate while the bremsstrahlung-Compton cooling
 is dominant for high mass accretion rates ($\dot M \gtrsim 10^{-3} {\dot
M}_{\rm Edd}$, $\Sigma \gtrsim 1.6 ~ {\rm g} ~ {\rm cm}^{-2}$) even
 though $\beta < 1$. The synchrotron cooling is relatively ineffective
 for higher mass accretion rates because of the lower 
electron temperature. As $\zeta$ increases, the electron temperature
 become high because the large magnetic flux enhances not only the
 synchrotron cooling but also the magnetic heating. As a result, the
synchrotron and synchrotron-Compton cooling become efficient. 
When $\zeta = 0.6$, 
the synchrotron-Compton is dominant in whole low-$\beta$ solutions.

Now we show that $\dot M \propto \Sigma$, $T \propto \Sigma^{-2}$,
$\beta \propto \Sigma^{-2}$ on the low-$\beta$ branch \citep[see
also][]{oda09}, where $T = (T_{\rm i} + T_{\rm e})/2$ is the mean
temperature. 
These relations depend on the dependence of ${\dot \Phi}_{\rm out}$ on
$\dot M$ (our outer boundary condition leads ${\dot \Phi} \propto {\dot
M}^{1/2}$). Here we introduce the parameter $s$, ${\dot \Phi}_{\rm out}
\propto {\dot M}^{s}$, in order to leave this dependence explicitly. 
First, we derive
the relations between $\dot M$ and $\Sigma$. Since $W_{\rm tot} \sim W_{\rm
mag} \propto T \Sigma \beta^{-1}$ on the low-$\beta$ branch, equations
(\ref{eq:h2}), (\ref{eq:mom_phi_int}), and (\ref{eq:al_int}) yield $H
\propto T^{1/2} \beta^{-1/2}$ and $\dot M \propto T \Sigma
\beta^{-1}$. Using equations (\ref{eq:con_int}), (\ref{eq:ind_int_ignore}),
and (\ref{eq:phidot}),  we find that $H \propto
\Sigma^{-(1-2s)/(7-4s)}$, $\beta \propto \Sigma ^{2(1-2s)/(7-4s)} T$,
and $\dot M \propto \Sigma^{1- 2(1-2s)/(7-4s)}$. 
Next, we derive the relation between $T$ and $\Sigma$ from the 
energy balance of the low-$\beta$ solutions ($Q^{+} \sim Q_{\rm
rad}^{-}$). 
Here we roughly approximate the
radiative cooling rate by $Q_{\rm rad}^{-} 
\propto \Sigma^{2} T^{1/2} H^{-1}$ for simplicity (this is the same
dependence as non-relativistic bremsstrahlung cooling for single temperature
plasma). Since $Q^{+} \propto W_{\rm tot} \propto \dot 
M$, we find that $T \propto \Sigma^{-2-6(1-2s)/(7-4s)}$. Therefore
$\beta \propto \Sigma ^{2(1-2s)/(7-4s)} T \propto
\Sigma^{-2-4(1-2s)/(7-4s)}$. We find that $\dot M
\propto \Sigma$, $T \propto \Sigma^{-2}$, and $\beta \propto
\Sigma^{-2}$ when $s=1/2$.

\begin{figure}
\epsscale{1.00}
\plotone{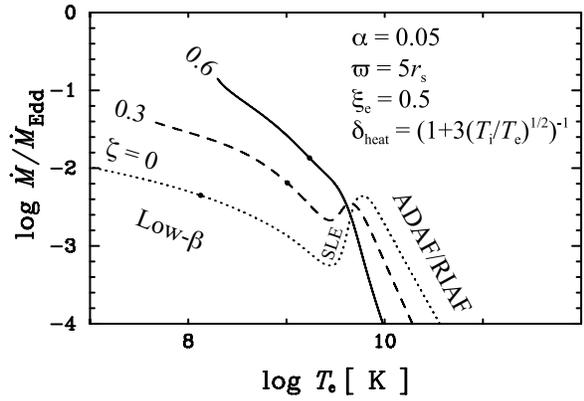}
\caption{Relations between the electron
temperature and the mass
 accretion rate for the same parameter as Figure
 \ref{al05ds3xp05sifoo}. Filled circles represent critical points at
 which $v_{\rm A} = \sqrt{c_{\rm s} v_{\rm K0}}$ (or equivalently $\beta
 = 2 c_{\rm s}/v_{\rm K0}$), where $v_{\rm A} = B_{\varphi} /\sqrt{4 \pi
 \rho_{0}}$ is the toroidal Alfv{\'e}n speed, $c_{\rm s}$ is the sound
 speed, and $v_{\rm K} = \varpi \Omega_{\rm K0}$ is the rotational
 speed. \label{al05ds3xp05temd}}
\end{figure}

We investigate the relation between the mass accretion rate and the
electron temperature in order to explain the anti-correlation between
the luminosity and the cutoff energy observed in the bright/hard
state. In the low-$\beta$ solutions, we
expect that the luminosity is roughly 
proportional to the mass accretion rate since $Q_{\rm rad}^{-} \sim
Q^{+} \propto \dot M$, and that the electron temperature roughly represents the
cutoff energy since the inverse-Compton scattering is the dominant radiative
cooling mechanism. 
Therefore, the relation between the electron temperature and the mass
accretion rate is useful for comparison with the observational data. 
Figure \ref{al05ds3xp05temd} shows the relations between
the electron temperature and the mass accretion rate for
the same parameters as in Figure \ref{al05ds3xp05sifoo}. 
We find that the electron temperature is typically $\lesssim 10^{9.5} ~ 
{\rm K}$ in the low-$\beta$ solutions while it is $\gtrsim 10^{9.5} ~  
{\rm K}$ in the ADAF/RIAF solutions. Furthermore, the electron
temperature in the low-$\beta$ solutions strongly anti-correlates
with the mass accretion rate.
Since $\dot M \propto \Sigma^{1- 2(1-2s)/(7-4s))}$ and $T \propto
\Sigma^{-2-6(1-2s)/(7-4s)}$ in the low-$\beta$ solutions, we find that
$T \propto {\dot M}^{-2 -2(1-2s)} \propto {\dot M}^{-2}$ for $s=1/2$.

According to \cite{pess05}, the MRI is stabilized for 
$v_{\rm A} \gtrsim \sqrt{c_{\rm s} v_{\rm K0}}$ (we plot this critical
point (filled circle in Figure \ref{al05ds3xp05temd}) at 
which $v_{\rm A} = \sqrt{c_{\rm s} v_{\rm K0}}$). Therefore, the
low-$\beta$ solutions may not exist under the condition that $v_{\rm A}
\gtrsim \sqrt{c_{\rm s} v_{\rm K0}}$. We
discuss this issue in Section \ref{discussion}.

\begin{figure}
\epsscale{1.0}
\plotone{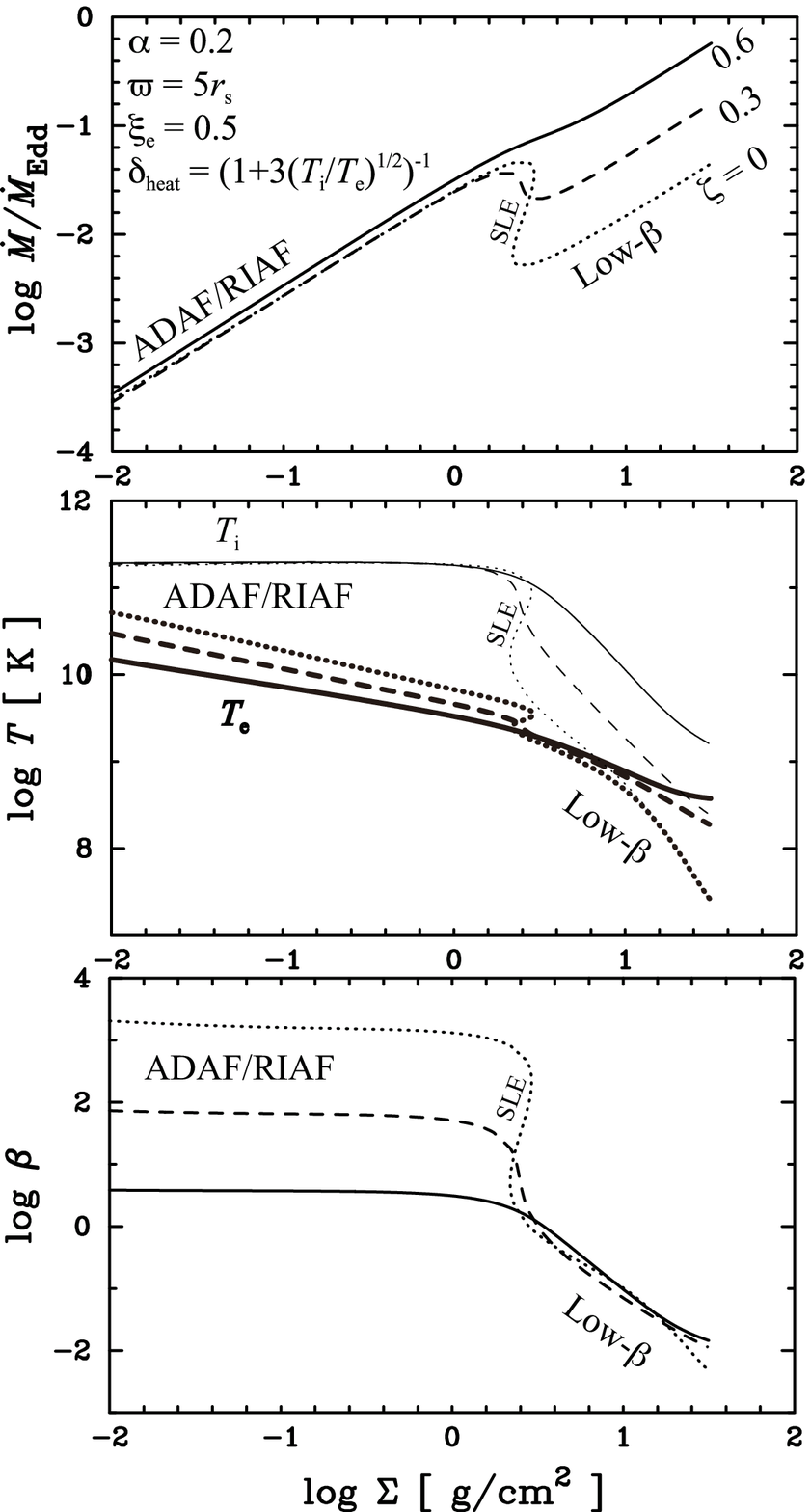}
\caption{Same as Figure \ref{al05ds3xp05sifoo} but for $\alpha =
 0.2$. \label{al20ds3xp05sifoo}}
\end{figure}

\begin{figure}
\epsscale{1.00}
\plotone{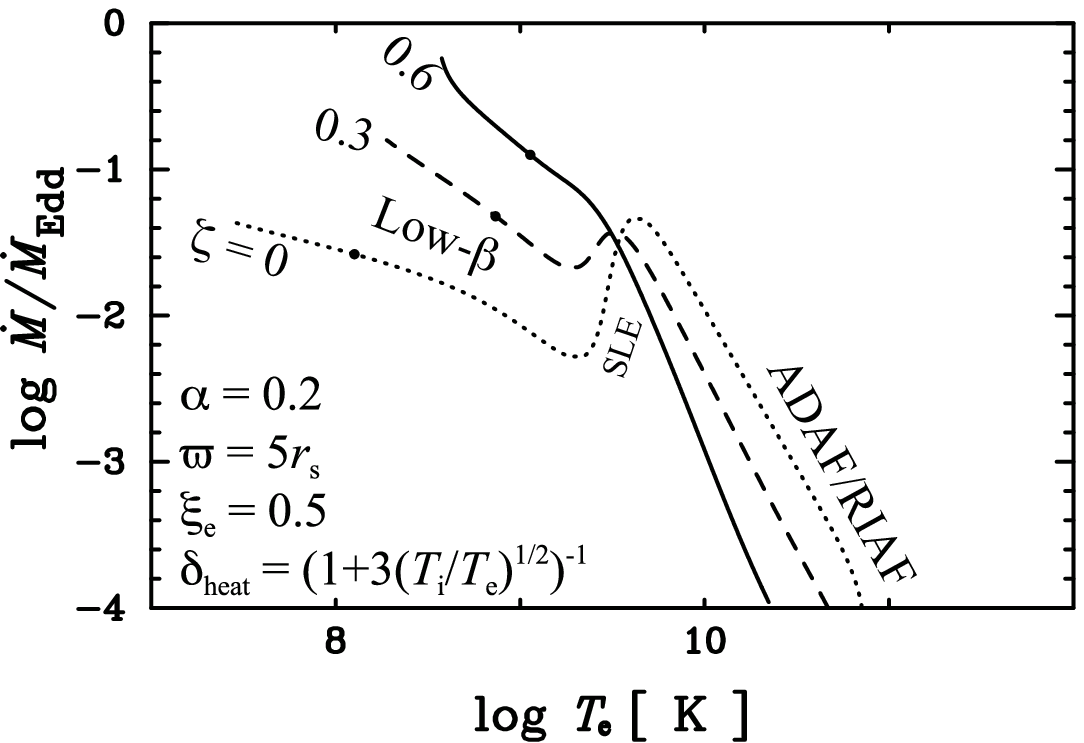}
\caption{Same as Figure \ref{al05ds3xp05temd} but for $\alpha =
 0.2$. \label{al20ds3xp05temd}}
\end{figure}

We also show the results for $\alpha = 0.2$ in Figure
\ref{al20ds3xp05sifoo} and \ref{al20ds3xp05temd}. The maximum
mass accretion rate of the ADAF/RIAF solution is around 
${\dot M}_{\rm c, A} 
\sim 0.05 {\dot M}_{\rm Edd}$. 
This maximum mass accretion rate is higher than that for $\alpha =
0.05$. 
Here we investigate the dependence of ${\dot
M}_{\rm c, A}$ on $\alpha$. 
Since $W_{\rm tot} \sim W_{\rm gas}$,
$Q_{\rm ad,i}/(1-\delta_{\rm heat}) Q^{+} \sim 
0.5$ and $Q_{\rm ad,e} \ll \delta_{\rm heat} Q^{+} + Q^{\rm ie}$ at $\dot M \sim
{\dot M}_{\rm c, A}$, the
energy equations yield (i) $Q_{\rm ad,i} \sim 0.5 (1-\delta_{\rm heat}) Q^{+}$, 
(ii) $Q^{\rm ie} \sim 0.5 (1-\delta_{\rm heat}) Q^{+}$, and (iii) $\delta_{\rm heat} Q^{+} +
Q^{\rm ie} \sim [\delta_{\rm heat} + 0.5 (1-\delta_{\rm heat})] Q^{+} \sim Q_{\rm
rad}^{-}$. 
Here we assume $\delta_{\rm heat}$ to be constant for simplicity. 
We find from Equation (i) that $T_{\rm i} \sim 10^{11} {\rm K}$. 
If $Q_{\rm rad}^{-}$ can be written in a simple form of $\Sigma^2/H \times
f(T_{\rm e})$ like $Q_{\rm br}^{-}$, we find from Equations (ii) and
(iii) that $T_{\rm e}$ at ${\dot M}_{\rm c,A}$ is independent of
$\alpha$ and ${\dot M}_{\rm c, A}$ 
is proportional to $\alpha^2 {\dot M}_{\rm Edd}$ exactly. Although
$Q_{\rm rad}^{-}$ has a complicated form in our model, our numerical
results indicate that the $T_{\rm e}$ is roughly 
independent of $\alpha$ ($T_{\rm e} \sim 10^{9.5} {\rm K}$). Using this
value of $T_{\rm e}$ instead of solving Equation (iii), we find from
Equation (ii) that roughly 
${\dot M}_{\rm c,A} \sim \alpha^2 {\dot M}_{\rm Edd}$. This result is
consistent with ${\dot M}_{\rm c,A} \sim 1.3 √\alpha^{2} {\dot M}_{\rm Edd}$ by
\cite{esin97}.

\subsection{Effect of  $\delta_{\rm heat}$}
We investigated the dependence of the results on $\delta_{\rm heat}$ because
$\delta_{\rm heat}$ is a poorly constrained parameter. We considered three cases:
$\delta_{\rm heat} = 0.5$ as an 
example of model in which ions and electrons receive equal amounts of
the dissipated 
magnetic energy, $\delta_{\rm heat} = 10^{-3}$ ($\sim m_{\rm e} / m_{\rm i}$) as an
example of a conventional model in which ions receive a substantial
amount of the energy, and $\delta_{\rm heat} = 0.2$ as an intermediate example. We
note that $\delta_{\rm heat} = (1+3\sqrt{T_{\rm
i}/T_{\rm e}})^{-1}$ lies between $\sim 0.05$ and $0.25$ in the solutions
presented in this paper.

Figure \ref{al05xp05sifoo} shows the thermal equilibrium curves for the
same parameters as in Figure \ref{al05ds3xp05sifoo} but for $\delta_{\rm heat} =
(1+3\sqrt{T_{\rm
i}/T_{\rm e}})^{-1}$ (solid), $0.5$ (dashed), $0.2$ (dotted), and
$10^{-3}$ (dot-dashed), respectively. The energy balance for ions and
electrons is illustrated in Figure \ref{al05xp05sieie}. 

The ion temperature decreases as $\delta_{\rm heat}$ increases because the
magnetic heating for ions ($(1-\delta_{\rm heat}) Q^{+}$) which is the
only heating source for ions decreases. ${\dot M}_{\rm c,A}$ also
decreases because of the decrease in the magnetic heating for ions. 
The electron temperature increases as
$\delta_{\rm heat}$ increases in the ADAF/RIAF solutions. By contrast,
in the low-$\beta$ solutions, the electron temperature is almost
independent of the value of $\delta_{\rm heat}$ because $\delta_{\rm
heat}$ does not appear practically in the energy equation for electrons
($\delta_{\rm heat} Q^{+} + Q^{\rm ie} \sim Q^{+} \sim Q_{\rm
rad}^{-}$). Therefore, we find that the exact value of $\delta_{\rm
heat}$ is not so important for the low-$\beta$ solutions.

\subsection{Dependence on the entropy gradient parameters, $\xi_{\rm i}$
  and $\xi_{\rm e}$}

Figure \ref{al05ds3xp05z000sifoo} shows the thermal equilibrium
curves for the same parameters as in Figure \ref{al05ds3xp05sifoo} but
for $\xi_{\rm i} = -1.0$ (dashed). We also plotted the thermal
equilibrium curves for $\xi_{\rm i} = 1.0$ (solid) for comparison. 

We obtained the LHAF solutions in the high mass accretion rate and high surface
density region. The heat advection works as a heating for
ions and balances the energy transfer via the Coulomb collisions in the
LHAF solutions. Above ${\dot M}_{\rm c,A}$, the heat advection term
overwhelms the magnetic heating. Hence, the ion temperature becomes
higher than that in the ADAF/RIAF solutions.

In the LHAF solutions, the electrons receive a
larger amount of heat from ions than that in the ADAF/RIAF
solutions. Nonetheless, the electron temperature
becomes lower than that in the ADAF/RIAF solutions (yet higher than that in the
low-$\beta$ solutions) because the radiative cooling becomes more
effective in such higher surface density region.

We find that our results are practically unchanged for any value of
$\xi_{\rm e}$ between $-0.5$ and $0.5$ because the heat advection term
is negligible
compared to the other terms in energy equation for electrons except when
$\delta_{\rm heat} = 10^{-3}$. Furthermore, even when $\delta_{\rm heat}
= 10^{-3}$, the other quantities except the electron temperature in the
ADAF/RIAF solutions does not change practically. We show the thermal equilibrium
curves plotted on the $\Sigma$ - $T_{\rm i}$ (thin) and $T_{\rm e}$
(thick) plane for $\delta_{\rm heat} = 10^{-3}$, $\xi_{\rm e} = 0.5$
(solid), $0$ (dashed), $-0.5$ (dotted), $\zeta = 0.6$ (left panel), and $0$
(right panel) in Figure \ref{al05dd3sitt}.

\section{Discussion} \label{discussion}

\subsection{Optically Thin, Magnetically Supported, Moderately Cool
  Disks} \label{discussion_main}
We obtained thermal equilibrium solutions for an optically thin,
two-temperature accretion disk incorporating magnetic fields. We
included bremsstrahlung emission, synchrotron emission, and inverse Compton
scattering as the radiative cooling mechanisms, and introduced the
parameter $\delta_{\rm heat}$ which represents the fraction of the magnetic heating
to electrons. We 
prescribed the $\varpi \varphi$-component of the azimuthally averaged
Maxwell stress is proportional to the total pressure. 
In order to complete the set of basic equations, we specified the radial
distribution of the magnetic
flux advection rate by introducing a parameter
$\zeta$. 
We found a branch of low-$\beta$ solutions in addition to the usual
ADAF/RIAF (for $\xi_{\rm i} > 0$), SLE, and LHAF (for $\xi_{\rm i} < 0$)
solutions.

Here we remark why we can obtain the low-$\beta$ solutions. First, we
prescribed that the $\varpi \varphi$-component of the Maxwell stress is
proportional to the total pressure. Therefore, if the magnetic pressure is
high, we can obtain the magnetic heating rate balancing the radiative
cooling rate even in high surface density and low temperature
region. Second, we specified the magnetic flux advection rate, $\dot
\Phi$, in order to complete the set of the basic
equations. In the conventional theory, $\beta$ is assumed to be constant
(typically, $\beta \sim 1$), which means
that the magnetic pressure is proportional to the gas pressure. This
implies that $p_{\rm gas}$ is just
multiplied by a constant ($(1 - \beta^{-1})p_{\rm gas}$). As a result,
we cannot obtain the sequence of the
low-$\beta$ solutions. On the other hand in our model, a decrease in
temperature results in an increase in magnetic pressure (therefore an
increase in the
magnetic heating) under the conservation of the magnetic flux in the vertical
direction. This is the reason why we can obtain the sequence of the
low-$\beta$ solutions.

We note that the exact values of $\xi_{\rm i}$, $\xi_{\rm e}$, and 
$\delta_{\rm heat}$ are not so important in the low-$\beta$ solutions. 
The magnetic heating enhanced by the high
magnetic pressure balances the radiative cooling in the low-$\beta$
solutions. The heat advection terms including $\xi_{\rm i}$ and $\xi_{\rm
e}$ are negligible for both ions
and electrons. Furthermore, $\delta_{\rm heat}$ does not appear in the energy
balance for electrons practically.

Let us discuss the lower limit of $\beta$. Since $\beta \propto
\Sigma^{-2}$ and $\Sigma \propto \dot M$ under our outer boundary
condition ($s=1/2$), we find that
$\beta \propto
{\dot M}^{-2}$, that is, $\beta$ decreases as the mass accretion rate increases
in the low-$\beta$ solutions.
It has been confirmed by global MHD simulations \citep[e.g.,][]{mach06}
that the MRI is not stabilized and hence the magnetic heating rate is
expressed in the form of equation  (\ref{eq:qmag}), at least, when
$\beta \gtrsim 0.1$. Local MHD simulations \cite[e.g.,][]{joha08} also
indicated that the Maxwell and Reynolds stresses generated by magnetic
turbulence are significant and yield an effective $\alpha$-viscosity
($\alpha \sim 0.1$) in highly magnetized disks ($\beta \sim 1$). However, the
expression of the magnetic heating employed in our paper may no longer
be valid for much lower values of $\beta$ because strong magnetic fields
suppress the growth of the MRI. We implicitly assumed that the
dissipation energy of the turbulent magnetic fields generated by the MRI
is converted into the thermal energy of the disk gas. Therefore, such
heating mechanism becomes ineffective if the MRI is stabilized. 
\cite{pess05} studied the evolution of the MRI in differentially rotating,
magnetized flows beyond the weak-field limit, and showed that the MRI is
stabilized for toroidal Alfv{\'e}n speeds, $v_{\rm A} = B_{\varphi} /
\sqrt{4 \pi \rho_{0}}$, exceeding the geometric mean of the sound speed,
$c_{\rm s}$, and the rotational speed, $v_{\rm K} = \varpi \Omega_{\rm
K0}$, (i.e., $v_{\rm A} \gtrsim \sqrt{c_{\rm s} v_{\rm K0}}$, or
equivalently $\beta \lesssim 2 c_{\rm s}/v_{\rm K0}$). 
Our results satisfy this condition when $\beta \lesssim 0.27 ~
(\zeta = 0.6)$, $0.09 ~ (\zeta = 0.3)$, 
$0.03 ~ (\zeta = 0)$ for $\alpha = 0.05$, and $\beta \lesssim 0.26 ~
(\zeta = 0.6)$, $0.08 ~ (\zeta = 0.3)$, and $0.02 ~ (\zeta = 0)$ for $\alpha
= 0.2$, respectively. These critical points are denoted by filled circles in
Figure \ref{al05ds3xp05temd} and \ref{al20ds3xp05temd} (see also
Table \ref{tbl-1}).
When $\beta$ falls below this critical value, the low-$\beta$ solutions
may not exist because there is no heating source balancing the radiative
cooling. As a result, the disk may undergo a transition to other states
(e.g, an MDAF-like disk or an optically thick disk).

\subsection{Thermal Stability}
Let us discuss the thermal stability in this subsection. A general
criterion concerning the thermal instability of
disks can be expressed as \citep[see][]{prin76,kato08}
\begin{eqnarray}
 \label{eq:criterion}
  \left[\frac{\partial}{\partial T} \left( -Q_{\rm ad} + Q^{+} - Q_{\rm
				     rad}^{-}\right)\right]_{\Sigma} >
  0 ~ .
\end{eqnarray}
In the low-$\beta$ solutions, we ignore the heat advection term 
because $Q^{+} \sim Q_{\rm rad}^{-} \gg Q_{\rm ad}$.
Once again we approximate the radiative cooling
rate by $Q_{\rm rad}^{-} \propto \Sigma^{2} T^{1/2} H^{-1}$ for
simplicity. 
We find from $Q^{+} \propto W_{\rm mag} \propto T \Sigma \beta^{-1} \propto
\Sigma^{1-2(1-2s)/(7-4s)}$ and $Q_{\rm rad}^{-} \propto \Sigma^{2}
T^{1/2} H^{-1} \propto \Sigma^{2+(1-2s)/(7-4s)} T^{1/2}$ (see
Section \ref{results_main}) that the
low-$\beta$ solutions do not satisfy the criterion 
(\ref{eq:criterion}), that is, are thermally stable. 
We note that the thermal stability is independent of $s$. 

We also remark the thermal stability of the other solutions (ADAF/RIAF,
SLE, and LHAF) in which the
gas pressure is dominant ($W_{\rm tot} \sim W_{\rm gas} \propto \Sigma
T$). Equations (\ref{eq:h2}) and (\ref{eq:con_int}) yield $H \propto
T^{1/2}$ and $\dot M \propto \Sigma T$. We find from $Q^{+} \propto
W_{\rm gas} \propto \Sigma T $, $Q_{\rm rad}^{-} 
\propto \Sigma^{2} T^{1/2} H^{-1} \propto \Sigma^{2}$, $Q_{\rm ad}
\propto \dot M T \xi \propto \Sigma T^{2} \xi$ that the ADAF/RIAF
solutions ($Q^{+} \sim Q_{\rm ad}$) are thermally stable but the SLE
($Q^{+} \sim Q_{\rm rad}^{-}$) and LHAF ($Q_{\rm ad} \sim Q_{\rm rad}^{-}$)
solutions are thermally unstable.

\subsection{A Candidate for the Bright/Hard State}
The main purpose of this paper is to explain the bright/hard state
observed during the bright/slow transition in
the rising phases of the transient outbursts of BHCs. In the low/hard state,
the X-ray spectrum is described by a
hard power law with a high energy cutoff at $\sim 200 ~ {\rm keV}$. 
When their luminosity exceed $\sim 0.1 L_{\rm Edd}$, these systems
undergo a transition from the low/hard state to the bright/hard state. In
the bright/hard state,
the cutoff energy decreases from $\sim 200 ~ {\rm keV}$ to $\sim 50 ~
{\rm keV}$ as the luminosity increases from $\sim 0.1 L_{\rm Edd}$ to
$\sim 0.3 L_{\rm Edd}$ \citep[e.g.,][]{miya08}. Beyond the bright/hard
state, these systems undergo a transition to the high/soft
state going through the VH/SPL state. 

The ADAF/RIAF solution explains 
the X-ray spectrum in the
low/hard state because the electron temperature is high ($T_{\rm e}
\gtrsim 10^{9.5} {\rm K}$). However, this solution
does not exist at the high mass accretion rates and does not show the
strong anti-correlation between the electron temperature and the mass
accretion rate observed in the bright/hard state. 
The low-$\beta$ solution extends to such high mass
accretion rates.
Therefore, the disk initially staying in the ADAF/RIAF state undergoes
transition to the low-$\beta$ state when the mass accretion rate exceeds ${\dot
M}_{\rm c, A}$. 
On the low-$\beta$ branches, the electron temperature
is low ($T_{\rm e} \sim 10^{8} - 10^{9.5} {\rm K}$ $\sim 10
- 300 ~ {\rm keV}$) and strongly anti-correlates with the mass accretion
rate. These features are consistent with the anti-correlation between the
luminosity and the energy cutoff observed in the bright/hard state.
Therefore, the low-$\beta$ solution can explain the bright/hard state.

In Section \ref{discussion_main}, we have discussed the lower limit
of $\beta$ below which the MRI is stabilized therefore the
low-$\beta$ solutions may not exist.
Since $\beta \propto {\dot M}^{-2}$ in the low-$\beta$ solutions, this
means that the low-$\beta$ solution has a maximum mass accretion rate,
${\dot M}_{{\rm c,} \beta }$. 
The mass accretion rate and the electron
temperature at the lower limit of $\beta$ are depicted by filled circles
in Figure \ref{al05ds3xp05temd} and \ref{al20ds3xp05temd} (see
also Table \ref{tbl-1}).
We found that when $\zeta \gtrsim 0.3$ (or equivalently, $\dot \Phi (\varpi = 5
r_{\rm s}) \gtrsim 4.9 {\dot \Phi}_{\rm out}$) 
${\dot M}_{{\rm c,} \beta} > {\dot M}_{\rm c, A}$. 
This indicates that 
the disk staying in the ADAF/RIAF state undergoes
transition to the 
low-$\beta$ disk when the mass accretion rate exceeds ${\dot M}_{\rm c,
A}$, after that, the disk undergoes transition to the optically thick
disk when the mass accretion rate exceeds ${\dot M}_{{\rm c,}
\beta}$. In this case, the hard-to-soft transition occurs at ${\dot M}_{{\rm c,}
\beta}$ (not at ${\dot M}_{\rm c, A}$). 
This can correspond to the bright/slow transition.

Here we also remark on the dark/fast transition during which the system
undergoes transition from the low/hard state to the high/soft state
without going through the bright/hard state and the VH/SPL state. 
We found that when $\zeta \lesssim 0.3$ (or equivalently, $\dot \Phi (\varpi = 5
r_{\rm s}) \lesssim 4.9 {\dot \Phi}_{\rm out}$) ${\dot M}_{{\rm c,}
\beta} < {\dot M}_{\rm c, A}$, that is, there is no
optically thin, thermally stable solution above ${\dot M}_{\rm c,
A}$. Therefore, the disk
staying in the ADAF/RIAF state immediately undergoes transition to an
optically thick disk without through the low-$\beta$ disk. This can
correspond to the dark/fast transition. 
Accordingly, we conclude that the bright/slow transition occurs
when $\dot \Phi$ has a large value and the
dark/fast transition occurs when $\dot \Phi$ has a small value.

\section{Summary} \label{summary}
We have obtained the low-$\beta$ solutions for optically thin,
two-temperature accretion disks incorporating the mean azimuthal
magnetic fields, and concluded that the low-$\beta$ solutions explain the
bright/hard state observed during the bright/slow transition of BHCs.
We assumed that the energy transfer from ions to electrons
occurs through Coulomb collisions, and considered bremsstrahlung
emission, synchrotron emission, and inverse Compton scattering as the
radiative cooling processes. We prescribed that the Maxwell stress is
proportional
to the total (gas and magnetic) pressure. In order to complete the set
of basic equations,
we specified the radial distribution of the magnetic flux advection
rate. Accordingly, a decrease in temperature can result in an increase in
magnetic pressure under the conservation of the magnetic flux in the vertical
direction. In the low-$\beta$ solutions, the magnetic heating is
enhanced by the high magnetic pressure. The fraction $(1-\delta_{\rm
heat})$ of the
magnetic heating goes into ions and is transferred to electrons via Coulomb
collisions. The fraction $\delta_{\rm heat}$ of the
magnetic heating goes into electrons. Eventually, the total amount of
the magnetic heating goes into electrons and balances the radiative
cooling (Compton cooling by bremsstrahlung and/or synchrotron photons). 
The electron temperature is lower than that in the ADAF/RIAF solutions
($T_{\rm e} \sim 10^{8} - 10^{9.5} {\rm K} \sim 10 - 300 ~ {\rm keV}$)
and strongly 
anti-correlates with the mass accretion rate in the low-$\beta$ solutions.
These features are consistent with the X-ray spectrum  observed in the
bright/hard state.

According to \cite{pess05}, the MRI is stabilized for $\beta
\lesssim 2 c_{\rm s}/v_{\rm K0}$. This indicates that the low-$\beta$
solutions disappear below this critical point. When the magnetic flux
advection rate is high ($\zeta \gtrsim 0.3$), the critical mass
accretion rate of the low-$\beta$ solutions, ${\dot M}_{{\rm c,}
\beta}$, is above 
the maximum mass accretion rate of the ADAF/RIAF solutions, ${\dot
M}_{\rm c, A}$. Therefore, the disk initially staying the ADAF/RIAF
state undergoes transition to the low-$\beta$ state when the mass
accretion rate exceeds ${\dot M}_{\rm c, A}$, after that, the disk
undergoes transition to the optically thick disk when the mass
accretion rate exceeds ${\dot M}_{{\rm c,}
\beta}$. This corresponds to the bright/slow transition. 
On the other hand, when the magnetic flux advection rate is low ($\zeta
\lesssim 0.3$), 
${\dot M}_{{\rm c,} \beta}$ is below ${\dot
M}_{\rm c, A}$. Therefore, the disk initially staying the ADAF/RIAF
state immediately undergoes transition to the optically thick disk when the mass
accretion rate exceeds ${\dot M}_{\rm c, A}$. This can correspond to
the dark/fast transition.

\acknowledgments
We are grateful to R. Narayan for valuable discussion and closely
examining my draft. This work is supported by the Grant-in-Aid for
Science Research of the Ministry of Education, Culture, Sports, Science
and Technology (RM: 20340040) and Grant-in-Aid for JSPS Fellows (20.1842).

\begin{onecolumn}
\begin{figure}
\epsscale{1.00}
\plotone{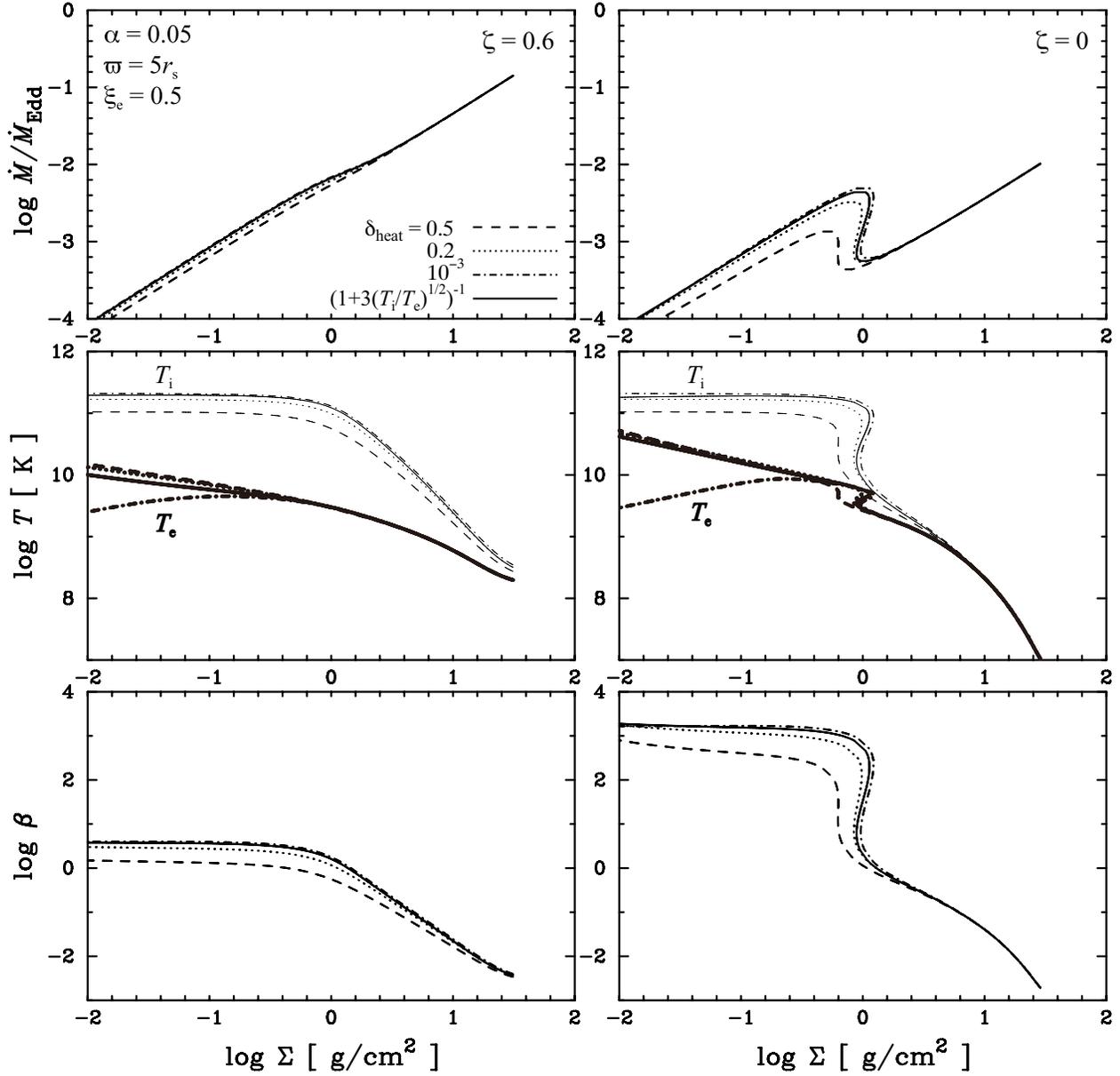}
\caption{Dependence of local thermal equilibrium curves on $\delta_{\rm heat}$. The
 disk parameters are $\alpha = 0.05$, $\xi_{\rm i} = 0.5$, $\xi_{\rm e}
 = 0.5$, $\delta_{\rm heat} = (1+3\sqrt{T_{\rm i}/T_{\rm e}})^{-1}$ (solid), $0.5$
 (dashed), $0.2$ (dotted), $10^{-3}$ (dot-dashed), $\zeta = 0.6$ (left)
 and $0$ (right). The exact value of $\delta_{\rm heat}$ is not so important for
 the low-$\beta$ solutions.  \label{al05xp05sifoo}}
\end{figure}

\begin{figure}
\epsscale{1.00}
\plotone{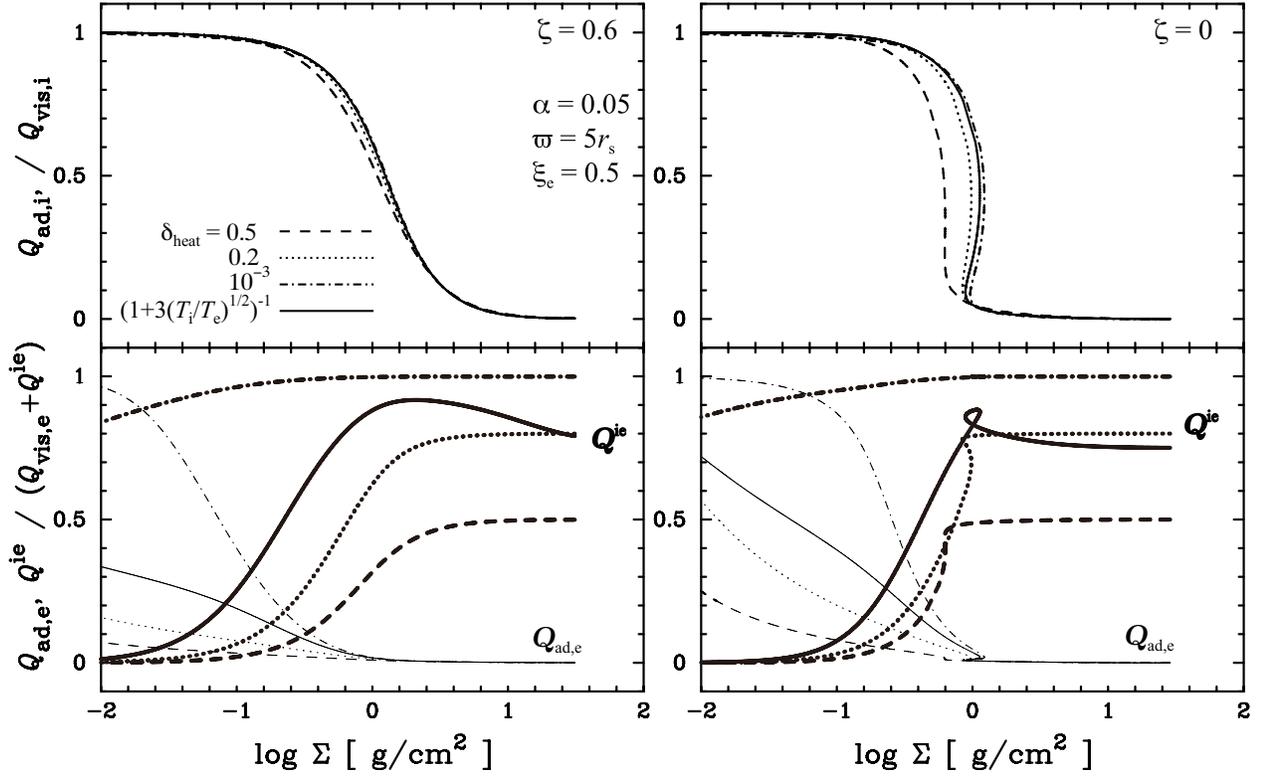}
\caption{Energy balance for ions and electrons for the same
 parameters as Figure \ref{al05xp05sifoo} (left: $\zeta = 0.6$, right:
 $\zeta = 0$). 
Upper: the ratio of the heat advection to the magnetic
heating for ions, $Q_{\rm ad,i}/(1-\delta_{\rm heat})Q^{+}$. 
Lower: the ratio of the heat advection to the total heating
 (thin), $Q_{\rm ad,e}/(\delta_{\rm heat} Q^{+} + Q^{\rm ie})$, and
 the fraction of the energy transfer via Coulomb collisions to the total
 heating (thick), $Q^{\rm ie}/(\delta_{\rm heat} Q^{+} + Q^{\rm ie})$, for
 electrons. \label{al05xp05sieie}}
\end{figure}
\end{onecolumn}

\begin{twocolumn}
\begin{figure}
\epsscale{1.0}
 \plotone{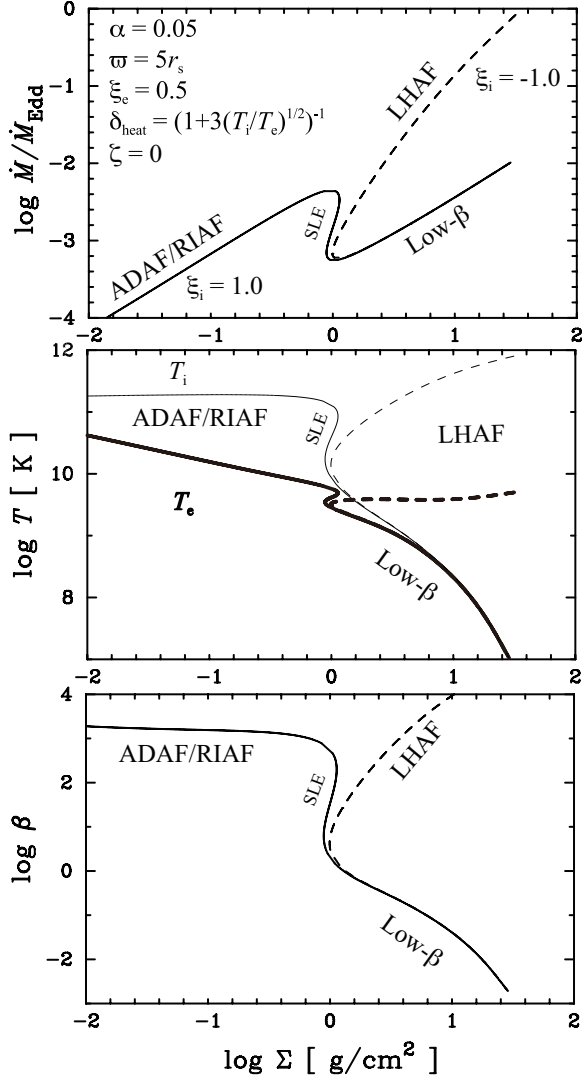}
 \caption{ Local thermal equilibrium
curves for the same parameters as in Figure \ref{al05ds3xp05sifoo} but
for $\xi_{\rm i} = -1.0$ (dashed). The equilibrium curves consist of the LHAF,
 SLE, and low-$\beta$ branches. We also plotted the thermal
equilibrium curves for $\xi_{\rm i} = 1.0$ (solid) for comparison. 
 \label{al05ds3xp05z000sifoo}}
\end{figure}
\end{twocolumn}

\begin{onecolumn}
\begin{figure}
\epsscale{1.00}
 \plotone{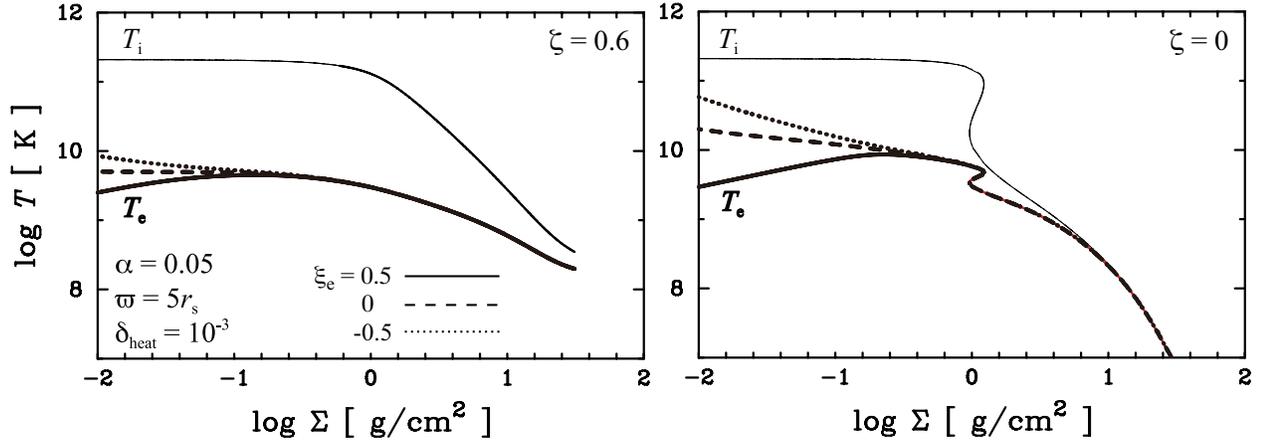}
 \caption{ Local thermal equilibrium curves plotted on the $\Sigma$ -
 $T_{\rm i}$ (thin) and $T_{\rm e}$ (thick) plane for $\delta_{\rm heat}=
 10^{-3}$, $\xi_{\rm e} = 0.5$ (solid), $0$ (dashed), $-0.5$ (dotted),
 $\zeta = 0.6$ (left panel), and $0$ (right panel) \label{al05dd3sitt}}
\end{figure}
\end{onecolumn}

\begin{onecolumn}
\begin{deluxetable}{llccc}
\tablecaption{Values at the lower limit of $\beta$ \label{tbl-1}}
\tablewidth{0pt}
\tablehead{
\colhead{$\alpha$} & \colhead{$\zeta$} & \colhead{$\beta$} & \colhead{${\dot M}_{{\rm c,} \beta}/{\dot M}_{\rm Edd}$} & \colhead{$T_{\rm
 e} ~  {\rm [ K ]}$} 
}
\startdata
$0.05$ & $0.6$ & $0.27$ & $1.3 \times 10^{-2}$ & $1.7
		 \times 10^9$ \\
 & $0.3$ & $0.09$ & $6.4 \times 10^{-3}$ & $1.0
		 \times 10^9$ \\
 & $0 $ & $0.03$  & $4.5 \times 10^{-3}$ & $1.3
		 \times 10^8$ \\
\tableline
$0.2$ & $0.6$ & $0.26$ & $1.2 \times 10^{-1}$ & $1.1
		 \times 10^9$ \\
 & $0.3$ & $0.08$ & $4.8 \times 10^{-2}$ & $7.4
		 \times 10^8$ \\
 & $0$ & $0.02$ & $2.6 \times 10^{-2}$ & $1.3
		 \times 10^8$ \\
\enddata
\end{deluxetable}

\end{onecolumn}

\end{document}